\documentclass[prd,aps,showpacs,preprintnumbers,amsmath,amssymb,superscriptaddress,twocolumn,floatfix]{revtex4}
\usepackage{alltt}
\usepackage{epsfig}
\usepackage{calc,rotating,xspace}
\usepackage{pifont}
\newcommand{\pt}{\ensuremath{p_{\rm T}}}
\newcommand{\met}  {\mbox{${\hbox{$E$\kern-0.6em\lower-.01ex\hbox{/}}}$}\ensuremath{_{\rm T}}}

\begin{document}
\clearpage

\title{\boldmath{Search for chargino-neutralino production in $\rm {p\bar p}$ collisions at $\sqrt s$ = 1.96 TeV with high-{\pt} leptons}}
\affiliation{Institute of Physics, Academia Sinica, Taipei, Taiwan 11529, Republic of China} 
\affiliation{Argonne National Laboratory, Argonne, Illinois 60439} 
\affiliation{Institut de Fisica d'Altes Energies, Universitat Autonoma de Barcelona, E-08193, Bellaterra (Barcelona), Spain} 
\affiliation{Baylor University, Waco, Texas  76798} 
\affiliation{Istituto Nazionale di Fisica Nucleare, University of Bologna, I-40127 Bologna, Italy} 
\affiliation{Brandeis University, Waltham, Massachusetts 02254} 
\affiliation{University of California, Davis, Davis, California  95616} 
\affiliation{University of California, Los Angeles, Los Angeles, California  90024} 
\affiliation{University of California, San Diego, La Jolla, California  92093} 
\affiliation{University of California, Santa Barbara, Santa Barbara, California 93106} 
\affiliation{Instituto de Fisica de Cantabria, CSIC-University of Cantabria, 39005 Santander, Spain} 
\affiliation{Carnegie Mellon University, Pittsburgh, PA  15213} 
\affiliation{Enrico Fermi Institute, University of Chicago, Chicago, Illinois 60637} 
\affiliation{Comenius University, 842 48 Bratislava, Slovakia; Institute of Experimental Physics, 040 01 Kosice, Slovakia} 
\affiliation{Joint Institute for Nuclear Research, RU-141980 Dubna, Russia} 
\affiliation{Duke University, Durham, North Carolina  27708} 
\affiliation{Fermi National Accelerator Laboratory, Batavia, Illinois 60510} 
\affiliation{University of Florida, Gainesville, Florida  32611} 
\affiliation{Laboratori Nazionali di Frascati, Istituto Nazionale di Fisica Nucleare, I-00044 Frascati, Italy} 
\affiliation{University of Geneva, CH-1211 Geneva 4, Switzerland} 
\affiliation{Glasgow University, Glasgow G12 8QQ, United Kingdom} 
\affiliation{Harvard University, Cambridge, Massachusetts 02138} 
\affiliation{Division of High Energy Physics, Department of Physics, University of Helsinki and Helsinki Institute of Physics, FIN-00014, Helsinki, Finland} 
\affiliation{University of Illinois, Urbana, Illinois 61801} 
\affiliation{The Johns Hopkins University, Baltimore, Maryland 21218} 
\affiliation{Institut f\"{u}r Experimentelle Kernphysik, Universit\"{a}t Karlsruhe, 76128 Karlsruhe, Germany} 
\affiliation{Center for High Energy Physics: Kyungpook National University, Daegu 702-701, Korea; Seoul National University, Seoul 151-742, Korea; Sungkyunkwan University, Suwon 440-746, Korea; Korea Institute of Science and Technology Information, Daejeon, 305-806, Korea; Chonnam National University, Gwangju, 500-757, Korea} 
\affiliation{Ernest Orlando Lawrence Berkeley National Laboratory, Berkeley, California 94720} 
\affiliation{University of Liverpool, Liverpool L69 7ZE, United Kingdom} 
\affiliation{University College London, London WC1E 6BT, United Kingdom} 
\affiliation{Centro de Investigaciones Energeticas Medioambientales y Tecnologicas, E-28040 Madrid, Spain} 
\affiliation{Massachusetts Institute of Technology, Cambridge, Massachusetts  02139} 
\affiliation{Institute of Particle Physics: McGill University, Montr\'{e}al, Canada H3A~2T8; and University of Toronto, Toronto, Canada M5S~1A7} 
\affiliation{University of Michigan, Ann Arbor, Michigan 48109} 
\affiliation{Michigan State University, East Lansing, Michigan  48824} 
\affiliation{University of New Mexico, Albuquerque, New Mexico 87131} 
\affiliation{Northwestern University, Evanston, Illinois  60208} 
\affiliation{The Ohio State University, Columbus, Ohio  43210} 
\affiliation{Okayama University, Okayama 700-8530, Japan} 
\affiliation{Osaka City University, Osaka 588, Japan} 
\affiliation{University of Oxford, Oxford OX1 3RH, United Kingdom} 
\affiliation{University of Padova, Istituto Nazionale di Fisica Nucleare, Sezione di Padova-Trento, I-35131 Padova, Italy} 
\affiliation{LPNHE, Universite Pierre et Marie Curie/IN2P3-CNRS, UMR7585, Paris, F-75252 France} 
\affiliation{University of Pennsylvania, Philadelphia, Pennsylvania 19104} 
\affiliation{Istituto Nazionale di Fisica Nucleare Pisa, Universities of Pisa, Siena and Scuola Normale Superiore, I-56127 Pisa, Italy} 
\affiliation{University of Pittsburgh, Pittsburgh, Pennsylvania 15260} 
\affiliation{Purdue University, West Lafayette, Indiana 47907} 
\affiliation{University of Rochester, Rochester, New York 14627} 
\affiliation{The Rockefeller University, New York, New York 10021} 
\affiliation{Istituto Nazionale di Fisica Nucleare, Sezione di Roma 1, University of Rome ``La Sapienza," I-00185 Roma, Italy} 
\affiliation{Rutgers University, Piscataway, New Jersey 08855} 
\affiliation{Texas A\&M University, College Station, Texas 77843} 
\affiliation{Istituto Nazionale di Fisica Nucleare, University of Trieste/\ Udine, Italy} 
\affiliation{University of Tsukuba, Tsukuba, Ibaraki 305, Japan} 
\affiliation{Tufts University, Medford, Massachusetts 02155} 
\affiliation{Waseda University, Tokyo 169, Japan} 
\affiliation{Wayne State University, Detroit, Michigan  48201} 
\affiliation{University of Wisconsin, Madison, Wisconsin 53706} 
\affiliation{Yale University, New Haven, Connecticut 06520} 
\author{T.~Aaltonen}
\affiliation{Division of High Energy Physics, Department of Physics, University of Helsinki and Helsinki Institute of Physics, FIN-00014, Helsinki, Finland}
\author{J.~Adelman}
\affiliation{Enrico Fermi Institute, University of Chicago, Chicago, Illinois 60637}
\author{T.~Akimoto}
\affiliation{University of Tsukuba, Tsukuba, Ibaraki 305, Japan}
\author{M.G.~Albrow}
\affiliation{Fermi National Accelerator Laboratory, Batavia, Illinois 60510}
\author{B.~\'{A}lvarez~Gonz\'{a}lez}
\affiliation{Instituto de Fisica de Cantabria, CSIC-University of Cantabria, 39005 Santander, Spain}
\author{S.~Amerio}
\affiliation{University of Padova, Istituto Nazionale di Fisica Nucleare, Sezione di Padova-Trento, I-35131 Padova, Italy}
\author{D.~Amidei}
\affiliation{University of Michigan, Ann Arbor, Michigan 48109}
\author{A.~Anastassov}
\affiliation{Rutgers University, Piscataway, New Jersey 08855}
\author{A.~Annovi}
\affiliation{Laboratori Nazionali di Frascati, Istituto Nazionale di Fisica Nucleare, I-00044 Frascati, Italy}
\author{J.~Antos}
\affiliation{Comenius University, 842 48 Bratislava, Slovakia; Institute of Experimental Physics, 040 01 Kosice, Slovakia}
% ...
\collaboration{CDF Collaboration\footnote{With visitors from $^a$University of Athens, 15784 Athens, Greece, 
$^b$Chinese Academy of Sciences, Beijing 100864, China, 
$^c$University of Bristol, Bristol BS8 1TL, United Kingdom, 
$^d$University Libre de Bruxelles, B-1050 Brussels, Belgium, 
$^e$University of California Irvine, Irvine, CA  92697, 
$^f$University of California Santa Cruz, Santa Cruz, CA  95064, 
$^g$Cornell University, Ithaca, NY  14853, 
$^h$University of Cyprus, Nicosia CY-1678, Cyprus, 
$^i$University College Dublin, Dublin 4, Ireland, 
$^j$University of Edinburgh, Edinburgh EH9 3JZ, United Kingdom, 
$^k$University of Heidelberg, D-69120 Heidelberg, Germany, 
$^l$Universidad Iberoamericana, Mexico D.F., Mexico, 
$^m$University of Manchester, Manchester M13 9PL, England, 
$^n$Nagasaki Institute of Applied Science, Nagasaki, Japan, 
$^o$University de Oviedo, E-33007 Oviedo, Spain, 
$^p$Queen Mary, University of London, London, E1 4NS, England, 
$^q$Texas Tech University, Lubbock, TX  79409, 
$^r$IFIC(CSIC-Universitat de Valencia), 46071 Valencia, Spain, 
}}
\noaffiliation

\date{\today}

\begin{abstract}
We present a search for the associated production of charginos
and neutralinos in $\rm {p \bar p}$ collisions at $\sqrt{s}$ = 1.96 TeV. The data were collected at the Collider Detector at Fermilab (CDF II)
and correspond to integrated luminosities between 0.7 and 1.0 fb$^{-1}$.
We look for final states with one high-\pt\  electron or muon, and two additional leptons.
Our results are consistent with the standard model expectations, and we
set limits on the cross section as a function of the chargino mass  in 
three different supersymmetric scenarios. For a specific MSSM scenario with no slepton mixing we 
set a 95\% C.L. limit at 151 GeV/$c^2$.
\end{abstract}

\maketitle
\newpage

\section{Introduction}
\label{intro}
\begin{figure}[h]
\begin{center}
\includegraphics[width=0.20\textwidth]{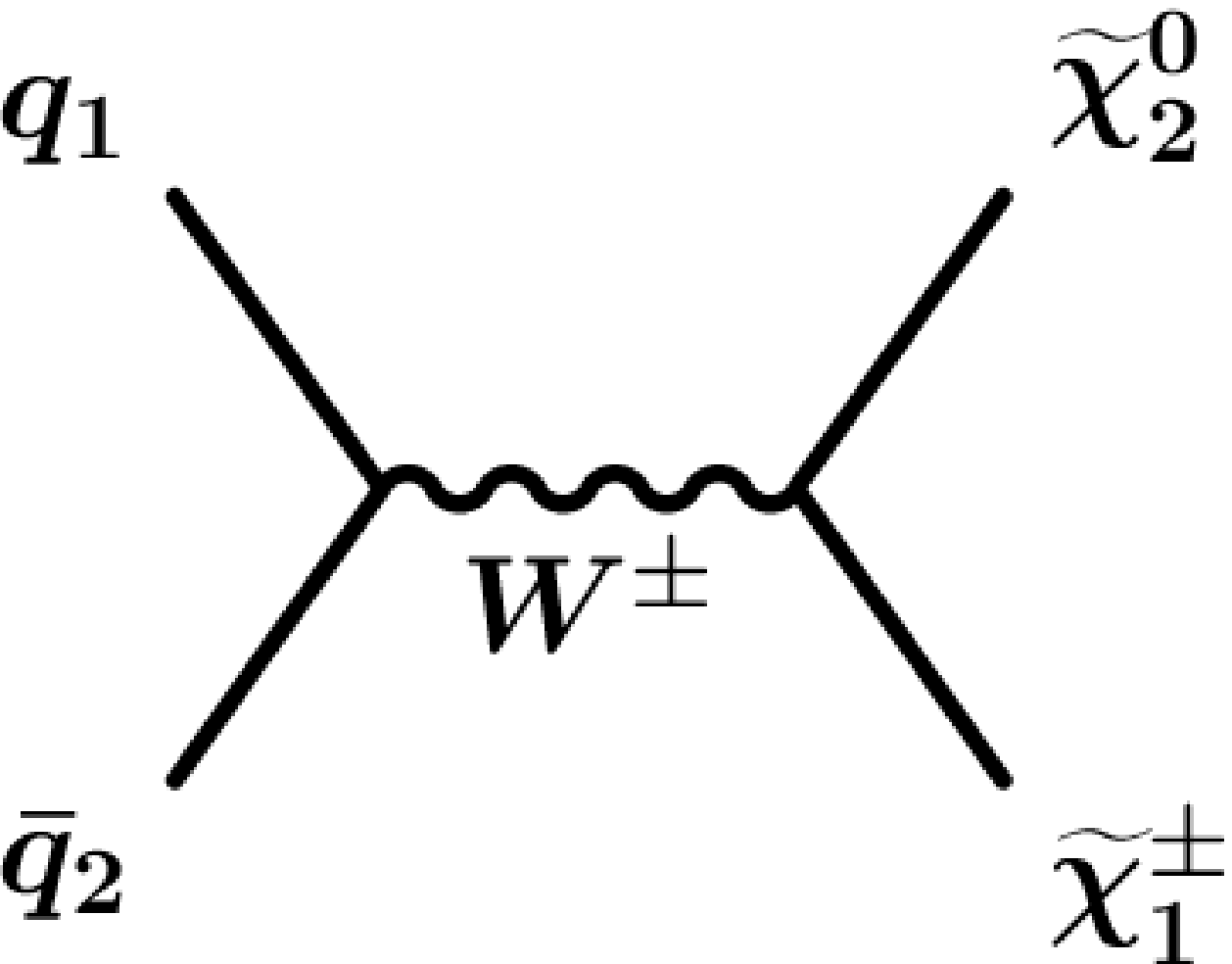}\hspace{0.6cm}
\includegraphics[width=0.20\textwidth]{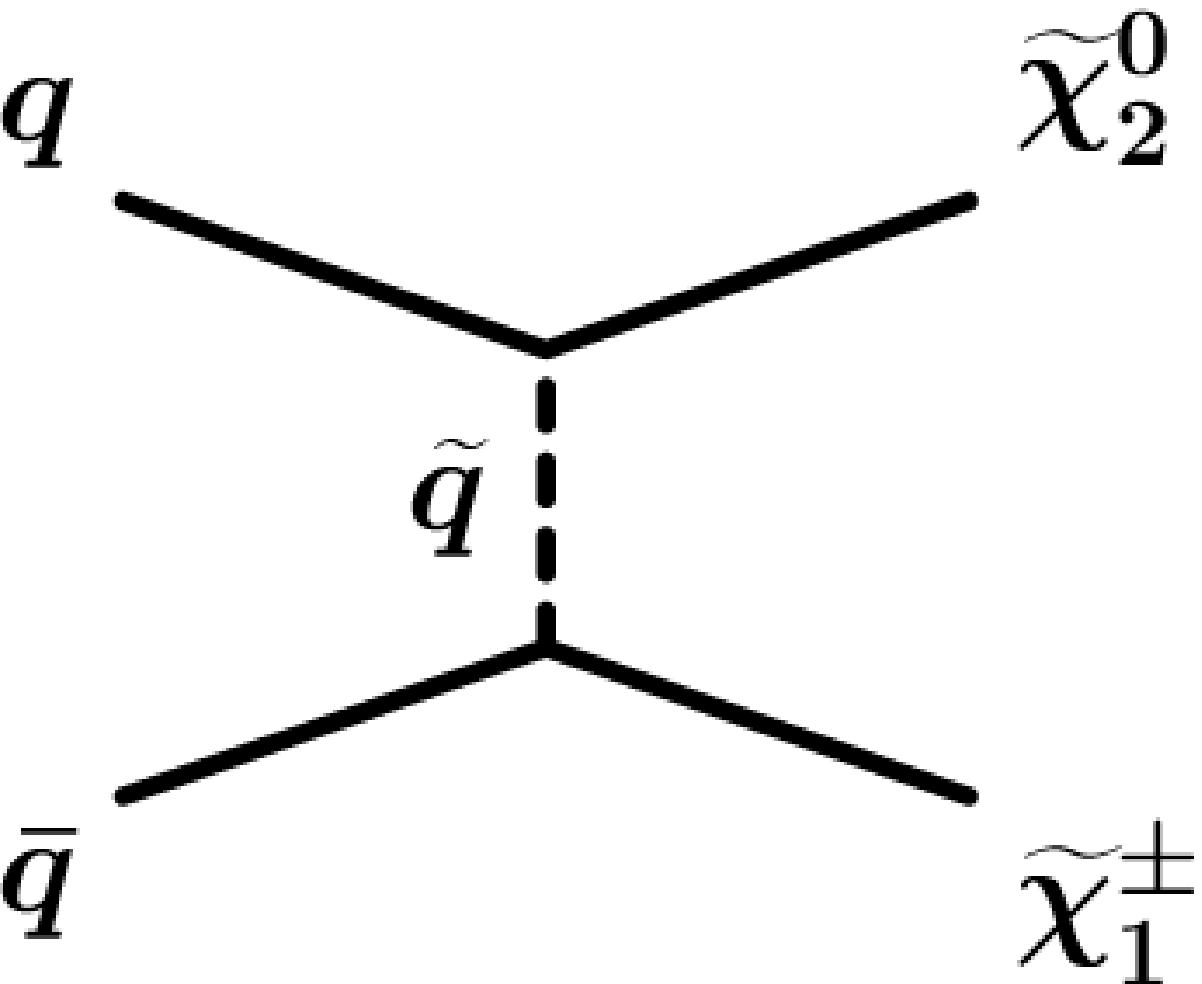}
\caption{Leading-order Feynman diagrams for chargino and neutralino associated production. The interaction is mediated through virtual W (left) and
squark ($\tilde{q}$, right). \label{fig::pro}}
\end{center}
\end{figure}

\begin{figure}[h]
\begin{center}
\includegraphics[width=0.20\textwidth]{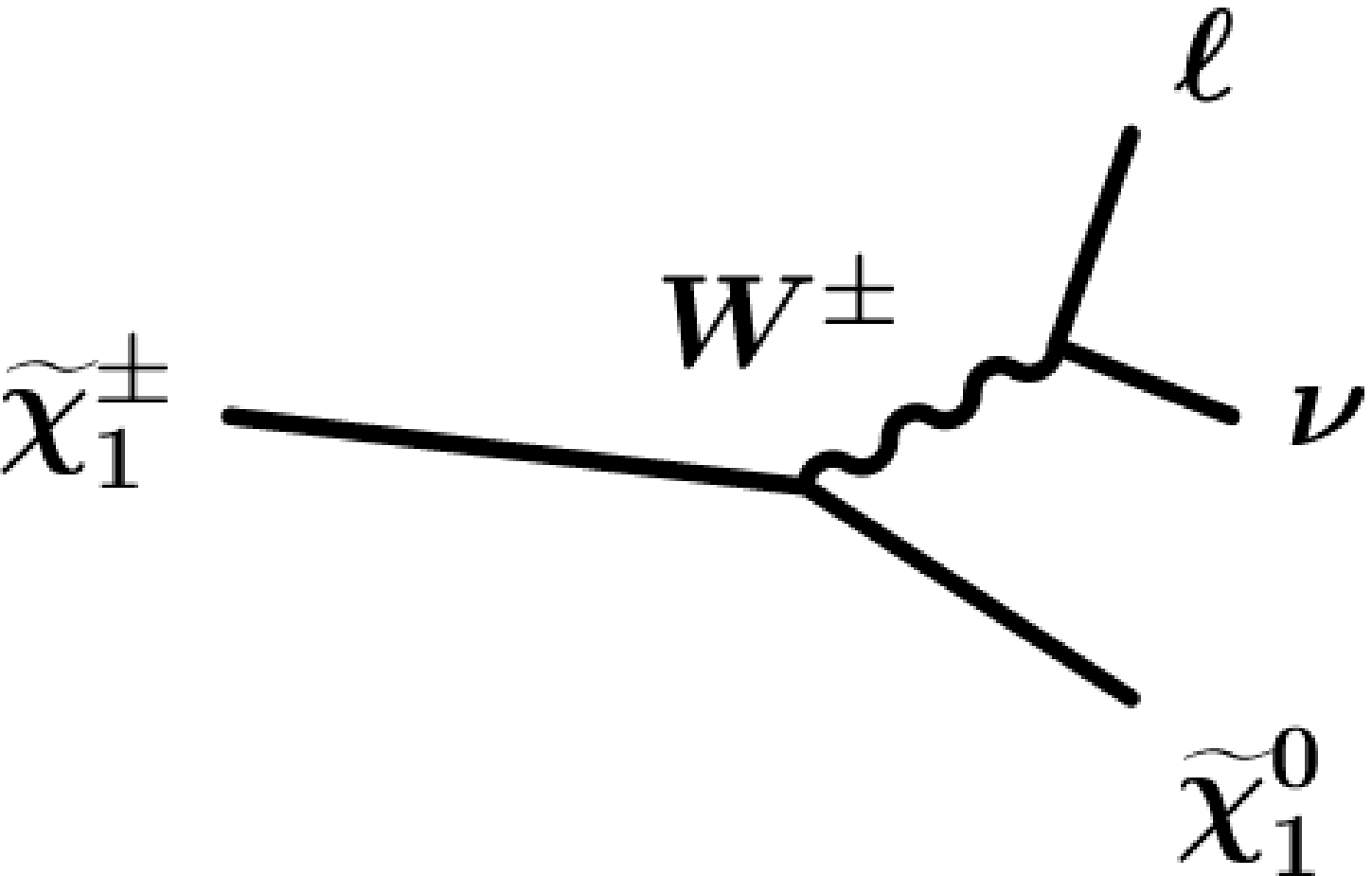}\hspace{0.5cm}
\includegraphics[width=0.20\textwidth]{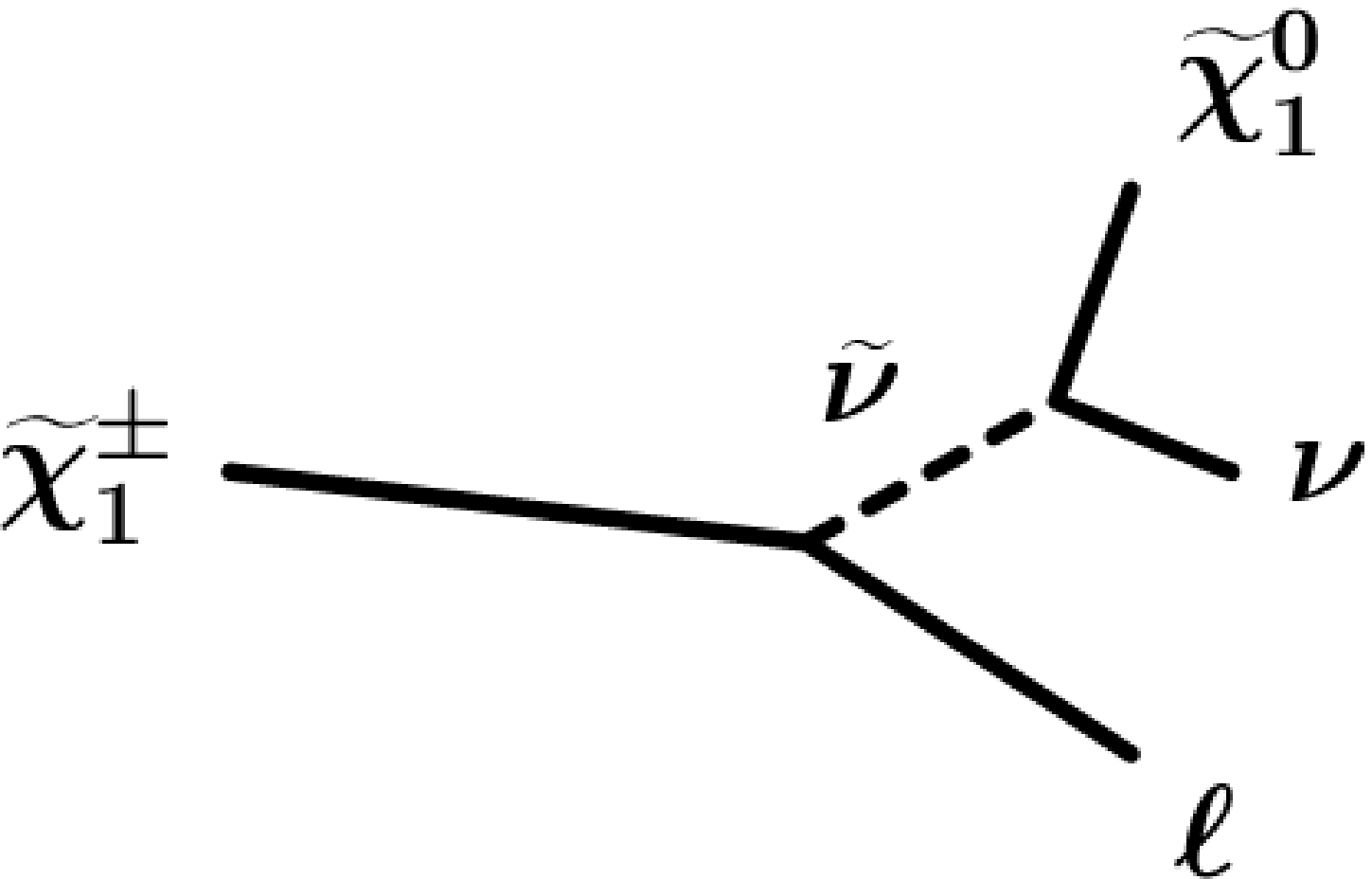}\vspace{0.5cm}
\includegraphics[width=0.20\textwidth]{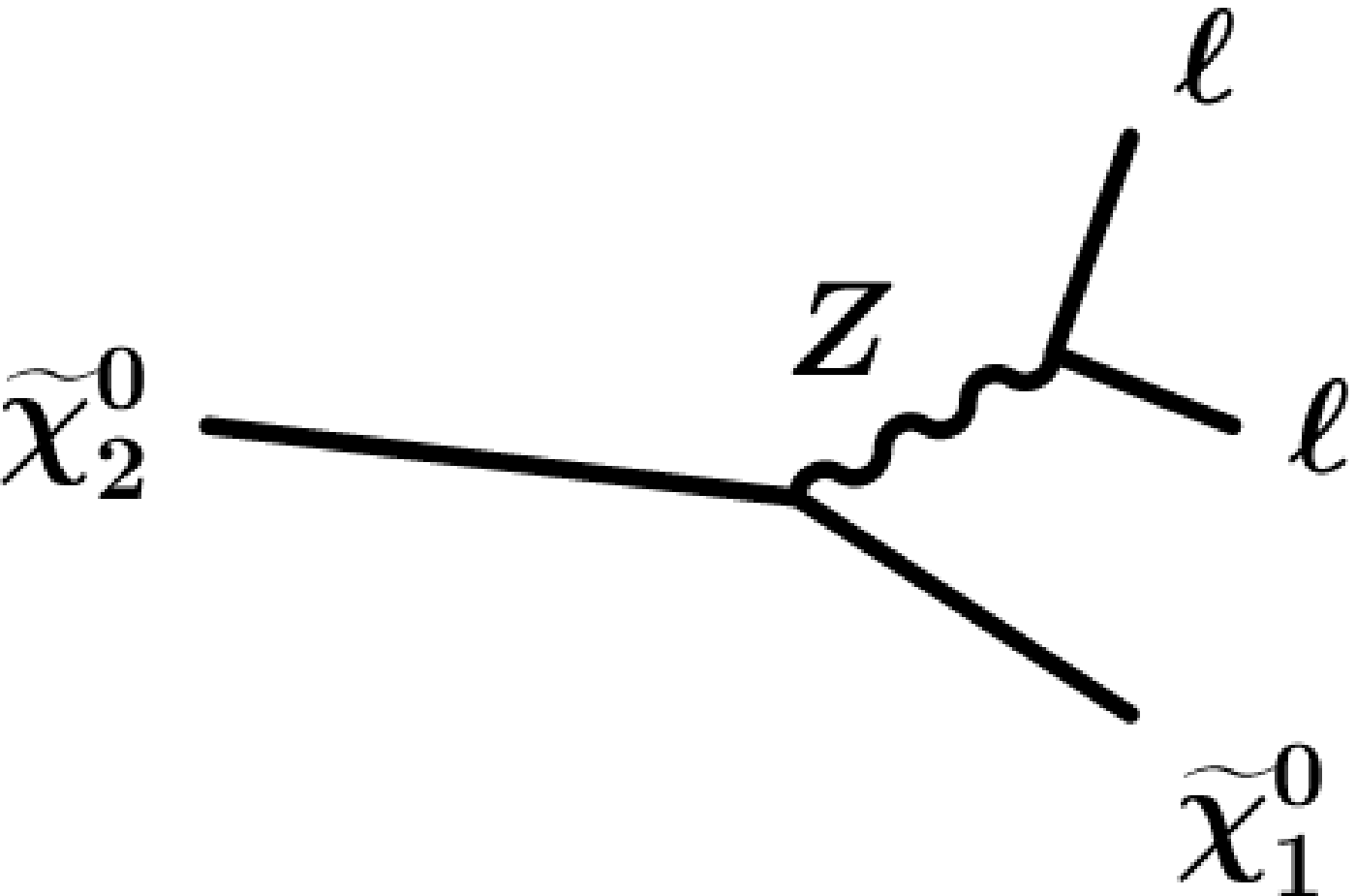}\hspace{0.5cm}
\includegraphics[width=0.20\textwidth]{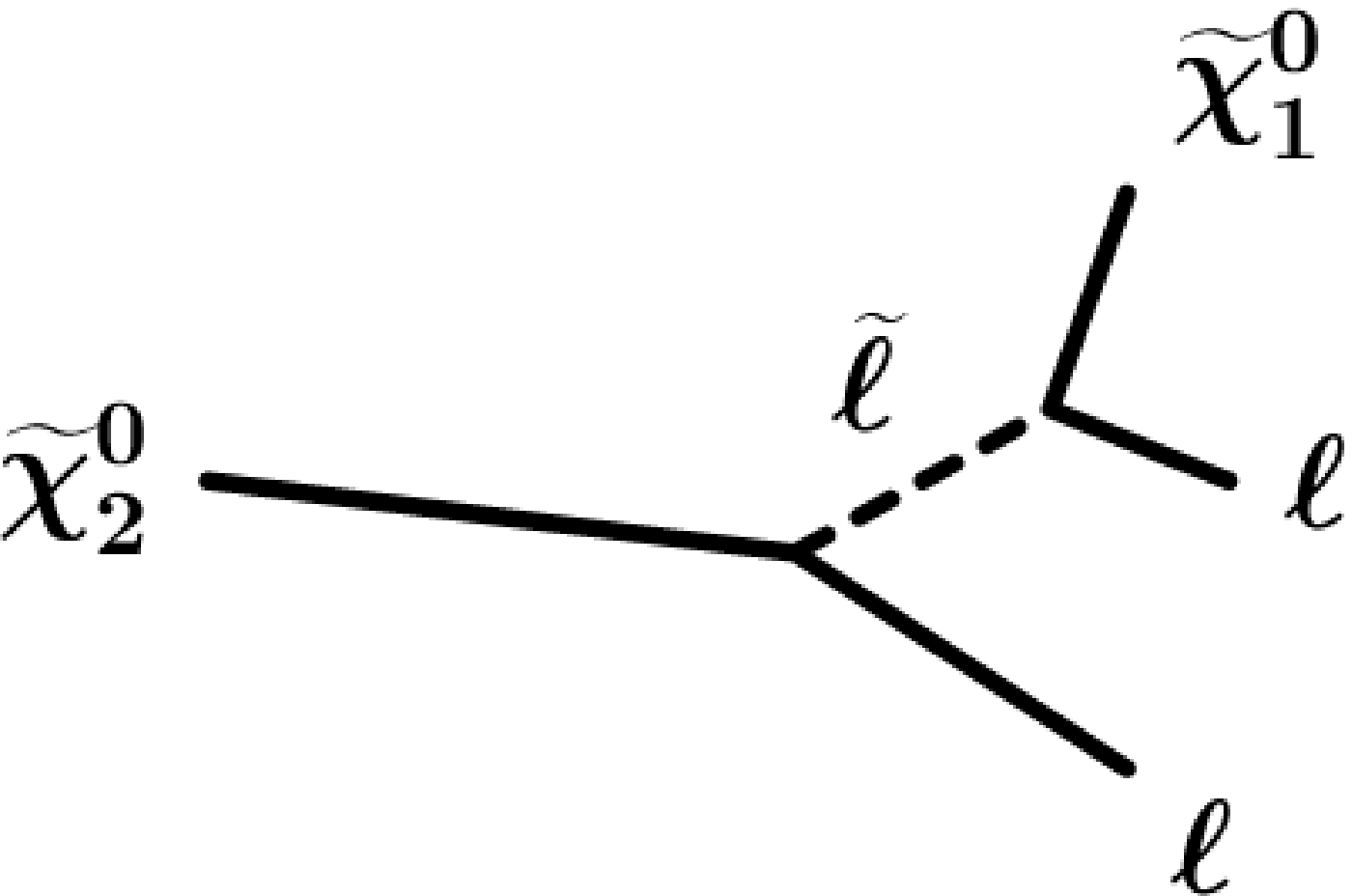}
\caption{Chargino and neutralino decay modes. The $\tilde{\ell}$ and the
$\tilde{\nu}$ are the SUSY counterparts of the lepton and the
neutrino. \label{fig::decay}}
\end{center}
\end{figure}

\indent Supersymmetry~\cite{prl1, martin} (SUSY) is a proposed symmetry of 
nature. 
It predicts the existence of supersymmetric partners for the standard 
model (SM) particles, called gauginos (higgsinos) for the gauge
(Higgs) bosons, and squarks/sleptons for fermions. The lightest SUSY particle is
referred to as the LSP.
If SUSY is an exact symmetry, the supersymmetric and the 
SM particles have the same mass, related couplings, and spin differing by 1/2.
As a consequence of the non-observation of light SUSY particles, such as the selectron,
SUSY must be a broken symmetry, if realized.
Several symmetry breaking models have been discussed in the past years.  
The gravitational interactions are responsible for the symmetry breaking in the mSUGRA~\cite{mSUGRA} scenario, whereas 
the ordinary gauge interactions are the source of SUSY breaking in the GMSB~\cite{martin} model.
In broken SUSY, 
gauginos and higgsinos combine to form mass eigenstates called charginos 
($\tilde{\chi}^{\pm}_{1,2}$) and neutralinos 
($\tilde{\chi}^{0}_{1, 2, 3, 4}$).
The lightest neutralino, $\tilde{\chi}^{0}_{1}$,
can be the LSP. 
SUSY is one of the most promising theories of physics beyond the 
SM as it can accommodate gravity
and unify the gauge interactions. 
In SUSY models where $R$-parity~\cite{rpv} is conserved, 
the LSP is stable and only weakly interacting, 
and thus is a viable  dark matter candidate. \\
\indent 
Experimental bounds on the gaugino masses are set by the LEP experiments
at 103.5 GeV/$c^2$ for the lightest chargino, in scenarios with large sfermion masses~\cite{LEP}, and
at 50.3 GeV/$c^2$ for the lightest neutralino in mSUGRA. 
These constraints are very robust within mSUGRA-inspired SUSY models and do not depend on the chargino decay modes, except for a few pathological cases~\cite{LEP2}.
The DO collaboration excludes the chargino mass below 117 GeV/$c^2$ in a specific SUSY breaking scenario described in~\cite{D0},
where the standard mixing between the left and the right components in the third generation families is suppressed. \\
\indent In this article we present a search for the associated production of 
the lightest chargino $\tilde{\chi}^{\pm}_{1}$ and the second-lightest neutralino $\tilde{\chi}^{0}_{2}$
(shown in Fig.~\ref{fig::pro}), performed as a counting experiment in data collected by the CDF detector.
Charginos and neutralinos can be among the lightest SUSY particles in the models we explore, with associated production 
cross sections within the reach of the Tevatron collider~\cite{carenaXS}.
If these sparticles decay leptonically within the detector, the final state is characterized
by the presence of leptons and significant missing energy~\cite{metdesc}, \met, due to particles escaping detection. 
While the process $\tilde{\chi}^{\pm}_{1}\tilde{\chi}^{0}_{1}\rightarrow \ell\nu\tilde{\chi}^{0}_{1}\tilde{\chi}^{0}_{1}$  results in a final state with $\ell+$\met, which has a large inclusive $W$ background, the distinct signature of 
$\tilde{\chi}^{\pm}_{1}\tilde{\chi}^{0}_{2}\rightarrow \ell\ell\ell\nu\tilde{\chi}^{0}_{1}$ 
makes the search for the associated production of chargino $\tilde{\chi}^{\pm}_{1}$ and
neutralino $\tilde{\chi}^{0}_{2}$ (see Fig.~\ref{fig::decay}), one of the most powerful tests of SUSY at hadron colliders.

\indent The paper is organized as follows. 
Section~\ref{cdf} contains a brief description of the CDF detector.
Section~\ref{leptonid} presents the lepton identification procedure and measurement of misidentification rate.
In  Section~\ref{dataSample} we discuss the backgrounds and in  Section~\ref{anaCuts} we 
describe the event selection.
In Section~\ref{sys} 
we present the estimated systematic uncertainties, followed by the validation of the analysis procedure in Section~\ref{cr}.
The searches presented in
this paper
are not targeted at a specific model, but rather are designed to cover a
large
range of possible new physics scenarios in which events with three leptons
and significant missing transverse energy are predicted at rates larger
than the SM predictions.
Nevertheless, based on the results presented in Section~\ref{results},
we set limits as a function of the chargino mass in several SUSY scenarios (Section~\ref{model}). The
results of the analysis presented in this paper are combined with results of similar searches
carried out at CDF to further improve the sensitivity~\cite{prl}.

\section{Experimental Apparatus}
\label{cdf}
The CDF II detector~\cite{cdf}
is a general-purpose detector with approximate azimuthal and forward-backward 
symmetry. CDF combines precision charged-particle tracking with projective
calorimeter towers and muon detection.
In the detector coordinate system, $\phi$ is the azimuthal angle around the beam axis and $\eta$ is the 
pseudorapidity defined as $\eta = - \ln \tan(\theta/2)$, where $\theta$ is the polar angle from the beam axis. 
The radial distance to the beam axis is referred to as $r$. \\
\indent The tracking system is composed of an inner
silicon detector ($1.5 < r < 29.0$ cm) and an outer drift chamber (COT, $40 < r < 140$ cm).
These detectors provide three-dimensional vertex measurement and  
track reconstruction efficiency  above 90\%
in the pseudorapidity range $|\eta|<2.0$.
For the leptons in our search, the resolution on the impact parameter is $\approx$ 40 $\mu$m, including
$\approx$ 30 $\mu$m for the beam size.
Surrounding the tracking system is a solenoidal
magnet which provides a 1.4~T field aligned parallel to the beam.
From the curvature of a track in the magnetic field, we determine the
transverse momentum \pt\ of charged particles. The momentum resolution of the outer tracking is $\sigma(p_{\rm T})/p_{\rm T}^2 = 0.0017$ $c$/GeV. \\
\indent Two layers of sampling calorimeters, one for detecting electromagnetic particles and the other to measure the
remaining hadronic energy, cover the range $|\eta|< 3.6$.
The central electromagnetic calorimeter (CEM) surrounds the solenoid within $|\eta|< 1.1$. It consists of lead sheets separated 
by polystyrene scintillator with an energy resolution of $13.5\%/\sqrt{E_{\rm T}} \oplus 1.5\%$, 
where $E_{\rm T} = |E|\sin\theta$ is measured in GeV. 
The CEM is segmented into 24 wedges per side; each wedge spans an angle of
approximately 15 degrees in $\phi$ and is divided into ten towers of $\Delta\eta=0.11$. At
normal incidence the total depth corresponds to about 18 radiation lengths ($X_0$). A proportional chamber (CES) is embedded in 
each CEM wedge at the shower maximum and provides good spatial resolution and shower shape information used for electron and photon identification. 
The central hadronic
calorimeter (CHA), positioned outside the CEM, matches the CEM segmentation into 24 wedges but 
uses steel absorbers interspersed with acrylic scintillators. There are
23 layers in the CHA and 
each layer is composed of one inch of steel and one
centimeter of scintillator.
The end wall 
calorimeter and the end plug calorimeter complete the coverage in the regions 
$0.8 < |\eta|< 1.2$ and $1.1 < |\eta|< 3.6$, respectively. The plug calorimeter
 consists of a lead-scintillator electromagnetic section (PEM) and an iron-scintillator hadronic section (PHA).  The PEM resolution is $16\%/\sqrt{E_{\rm T}} \oplus 1\%$.
The PEM also contains a shower maximum detector (PES). \\
\indent The muon system is installed outside the calorimeters. The innermost four-layer
drift chamber system (CMU) can detect minimum ionizing particles with transverse momenta larger than 1.4 GeV/$c$. An additional four-layer drift chamber (CMP) is located
outside the magnet return yoke and detects particles with \pt\ $ > $ 2.0 GeV/$c$. The 
CMU-CMP coverage ($|\eta|< 0.6$) is extended up to $|\eta|< 1.0$ by the central 
muon extension chambers (CMX). Outside the CMP and CMX chambers are scintillator detectors providing additional timing measurements. The last set of muon detectors (IMU) covers the region $1.0 < |\eta|< 1.5$. 
The information from the IMU chambers is not used in this analysis. \\ 
\indent The luminosity is measured from the total inelastic $\rm{p}\bar{\rm{p}}$ cross-section 
using Cherenkov counters located in the $3.7 <|\eta|< 4.7$ region.\\
\indent The CDF trigger has a three-level architecture.
The first level (L1) is a custom-designed hardware trigger 
which  makes a fast trigger decision based on
preliminary information from the tracking, calorimeter, and muon systems with
an average accept rate of 25 kHz.
The second level (L2) uses both custom hardware and a software-based event reconstruction
with an accept rate of 750 Hz. The third level (L3) uses the offline
reconstruction  software and selects events for storage with a rate of up to 85 Hz~\cite{runIIb}.

\section{Lepton Identification}
\label{leptonid}
\subsection{Lepton identification probability}
We use different constraints on identification variables for high-\pt\ (\pt\ $>$ 20 GeV/$c$) and low-\pt\ (\pt\ $<$ 20 GeV/$c$) leptons due to  different detection characteristics  and   also due to the trigger requirements. These identification criteria are described below and are summarized in Tables~\ref{tab:tracks} through~\ref{tab:muons}.\\
\indent Reconstructed central tracks must have at least  five hits out of 12 possible in at least 
three (two) out of four axial (stereo) COT super layers, to ensure high reconstruction efficiency and purity. We accept only tracks originating within $\pm$60 cm from the center of the detector, 
 and we apply a cut on the impact parameter ($d_0$, see Table~\ref{tab:tracks}) to suppress cosmic rays and secondary vertices. The impact parameter 
is the radial distance of closest approach between the track and the beam line. 
For each beam-constrained COT track, we place a requirement  on the fit quality 
 $\chi^2$ normalized by the number of degrees of freedom in the track fitting.
The efficiency of reconstructing a track is measured separately in calorimeter triggered  $W \rightarrow e \nu$  events as described in~\cite{trackeff}.\\ 
\indent A candidate electron in the central region is a track pointing to an electromagnetic calorimeter cluster. 
If the ratio of the energy measured in the hadronic calorimeter to that measured in the electromagnetic calorimeter is small,
we define it as a ``loose'' electron. 
%
Additional requirements
on the shower shape and the energy to momentum ratio are imposed
 to select high purity, ``tight'' electrons. One such requirement, the lateral shower sharing profile ($L_{\rm{shr}}$) compares the energy sharing between neighboring CEM towers to the expectation from test beam data. We also restrict the matching between the shower and the track, both the distance in the $r$-$\phi$ plane (Q$\cdot \Delta x$, where $Q$ is the lepton charge), and in the r-z plane ($|\Delta z|$). In addition we also restrict the $\chi^2$ of the fit to the shower profile in the CES, and to test beam data. A similar procedure based on the $\chi^2$ from comparing the tower energy distribution is applied to
  electrons reconstructed in the plug calorimeter. In this case only tracks within $|\eta|<2$ and with silicon hits are accepted. The collimation of the shower shape in the PES is also restricted, by requiring that the energy in the middle five strips of a PES cluster should be more than 65\% of the energy in all nine strips. 
If the track associated to the candidate electron is consistent with coming from a $\gamma \rightarrow e^+ e^-$ conversion, the candidate electron
is rejected. The photon conversion identification algorithm defines an electron as originating from a conversion if the
 azimuthal separation of the electron candidate and any oppositely-charged track 
at the tangency point ($D_{xy}=R\times\Delta\phi$ where $R$ is the conversion radius) is less than 
 0.2 cm 
and the difference in polar angle ($\Delta\cot\theta$)
 is smaller than 0.04. The measurement of the conversion identification efficiency is described in Section~\ref{conversion}. \\
\indent Tracks with small energy deposits in the calorimeters and matched stubs~\cite{stubfootnote} in the CMU and CMP (or CMX only) muon chambers are candidates for the CMUP (CMX)  muon category. The matching between the extrapolated track and the 
stub in the chamber ($|\Delta x|$, where $x$ is the local linear coordinate in the transverse plane) 
has to be within a certain range (refer to Table~\ref{tab:muons}). If a track has \pt\ less than 20 GeV/$c$, the effect of multiple scattering is enhanced and thus we set a less stringent requirement.
For CMX muons we restrict our selection to tracks that pass through all eight super layers of the COT. The efficiency of finding a stub in the first place is measured separately and combined with the other identification measurements.\\
\indent 
Other muons in an event are also included if they fall in the muon category called 
``central minimum ionizing objects'' (CMIO's). This category is composed of tracks with \pt\ greater than 10 GeV/$c$ for which the track does not
extrapolate to the fiducial region of the CMU and CMP or CMX chambers~\cite{cmio}.
In this case we constrain the selection to muon
candidate tracks with a non-zero calorimeter energy deposit to suppress tracks entering uninstrumented parts.
This extends the muon coverage to $|\eta|<1.5$, with lower efficiency and lower purity for $|\eta|>1.2$.\\
\indent Since leptons from $\tilde\chi_1^{\pm}$ and $\tilde\chi_2^0$ decays are
expected to be well separated from each other and from other objects in the event,~\cite{tevIIreport},
we restrict our studies to isolated electrons and muons. To decide whether a lepton is isolated or not we sum up the calorimeter transverse energy ($E_{\mathrm{T}}^{\mathrm{cone}}$) in a cone of $\Delta R = \sqrt{\Delta \eta^2 + \Delta \phi^2}<$ 0.4 around, but not including, the energy deposited by the lepton. 
We require  $E_{\mathrm{T}}^{\mathrm{cone}}$ to be less than 2 GeV for the loose CMUP and CMX muons.
The $E_{\mathrm{T}}^{\mathrm{cone}}$ is required to be smaller 
than  10\% of the muon {\pt} (electron $E_{\rm T}$) for other lepton categories, or if the muon~\pt\ (electron $E_{\rm T}$) is above 20 GeV/$c$ (GeV). 
\begin{table}[h]
\begin{center}
\caption{Requirements for central tracks}
\begin{tabular}{lr} \hline  \hline
Variable & Cut \\ \hline
no. axial COT super layers & \hspace{1cm} $\geq$ 3 with $\geq$ 5 hits\\ 
no. stereo COT super layers & $\geq$ 2 with $\geq$ 5 hits\\ 
$|z_0|$ & $<$ 60 cm \\
$|d_0|$ (no silicon hits)& $<$ 0.2 cm \\
$|d_0|$ (silicon hits)& $<$ 0.02 cm \\ \hline
Muon tracks: & \\
COT exit radius (CMX) & $>$ 140 cm \\
$\chi^2$ (first 350 pb$^{-1}$) & $<$ 2.8  \\
$\chi^2$ (otherwise) & $<$ 2.3 \\ \hline\hline
\end{tabular}
\label{tab:tracks}
\end{center}
\end{table}

\begin{table}[h]
\begin{center}
\caption{Electron selection criteria}
\begin{tabular}{lr} \hline  \hline
Variable & Cut \\ \hline
\bf{Loose electron:} & \\
Central track & \\
Not conversion & \\
$|\eta|$ & $<$ 1 \\
$E_{\rm {Had}}$/$E_{\rm {EM}}$ & $<$ 0.055+0.00045$\cdot$ $E_{\rm{EM}}$ GeV\\ 
$E_{\rm{T}}^{cone}$/$E_{\rm T}$ & $<$ 0.1 \\ \hline
\bf{Tight electron:} & \\
As electron1 except & \\
$E/p$ ($p_{\rm{T}} <$ 50 GeV/c) & $<$ 2 \\
$L_{\rm {shr}}$ & $<$ 0.2 \\
$\chi^2_{strips}$ & $<$ 10 \\
$Q\cdot \Delta x $ & $>$ -3.0,  $<$ 1.5 cm \\
$| \Delta z |$ & $<$ 3.0 cm \\ \hline
\bf{Plug electron:} & \\
Track with Silicon hits& \\
Not conversion & \\
$|\eta|$ & $>$ 1.2 $<$ 2.0 \\
$E_{\rm{Had}}$/$E_{\rm{EM}}$ & $<$ 0.055 GeV\\
$E_{\rm{T}}^{\rm {cone}}$/$E_{\rm{T}}$ & $<$ 0.1 \\ 
$E/p$  & $<$ 3 \\
PES 5/9 (see text) & $>$ 0.65 \\
$\chi^2_{\rm{PEM}}$ & $<$ 10 \\ \hline \hline
\end{tabular}
\label{tab:electrons}
\end{center}
\end{table}

\begin{table}[h]
\begin{center}
\caption{Muon selection criteria}
\begin{tabular}{lr} \hline  \hline
Variable & Cut \\ \hline
\bf{Base muon:} & \\
Central track & \\ 
$E_{\rm{EM}}$ & $<$ 2 GeV\\
$E_{\rm{Had}}$ ($p_{\rm{T}} \geq$ 20 GeV/c) & $<$ 6 GeV  \\
$E_{\rm{Had}}$ (p$_T <$ 20 GeV/c) & $<$ 3.5 + p$_T$/8 GeV  \\ \hline \hline
\bf{Tight CMUP/CMX muon:} & \\
 Base muon & \\
$E_{\rm{T}}^{\rm{cone}}$/$p_{\rm{T}}$ & $<$ 0.1 \\
$|\Delta x_{CMU}|$ (CMUP) & $<$ 7 cm or $\chi^2<$ 9\\
$|\Delta x_{CMP}|$ (CMUP) & $<$ 5 cm or $\chi^2<$ 9\\
$|\Delta x_{CMX}|$ (CMX) & $<$ 6 cm or $\chi^2<$ 9\\ \hline
\bf{Loose CMUP/CMX muon:} & \\
As muon1 except: & \\
$E_{\rm{T}}^{\rm{cone}}$ ($p_{\rm{T}} <$ 20 GeV/c) & $<$ 2 GeV  \\ \hline

\bf{CMIO muon:} & \\
Base muon & \\
Not fiducial to CMUP, CMX& \\
$p_{\rm{T}}$ & $\geq$ 10 GeV/$c$  \\
$E_{\rm{T}}^{\rm{cone}}$/$p_{\rm{T}}$ & $<$ 0.1 \\
$E_{\rm{EM}}$ + $E_{\rm{Had}}$ & $>$ 0.1 GeV\\ \hline \hline
\end{tabular}
\label{tab:muons}
\end{center}
\end{table}
\indent The primary data sample used to measure trigger and identification efficiencies for leptons with \pt\ ($E_{\rm{T}}$) $>$ 20 GeV/$c$ (GeV) is the single-electron or single-muon triggered sample with a \pt\ ($E_{\rm{T}}$) threshold of 18 GeV/c (GeV) used for the analysis itself, as described in Section~\ref{dataSample}. 
Samples of simulated events needed in this study are also presented in 
Section~\ref{dataSample}.
  Additional data samples are used to measure 
    efficiencies and misidentification probabilities for lower-energy
    leptons. These include samples collected with single-lepton
    trigger thresholds of 8 GeV/$c$ for muon \pt\ and 8 GeV for electron $E_{\rm{T}}$, 
    and inclusive central jet samples collected
    with jet trigger thresholds at $E_{\rm{T}} >$ 20, 50, 70, and 100 GeV.

\begin{figure}\begin{center}\includegraphics[width=0.45\textwidth]{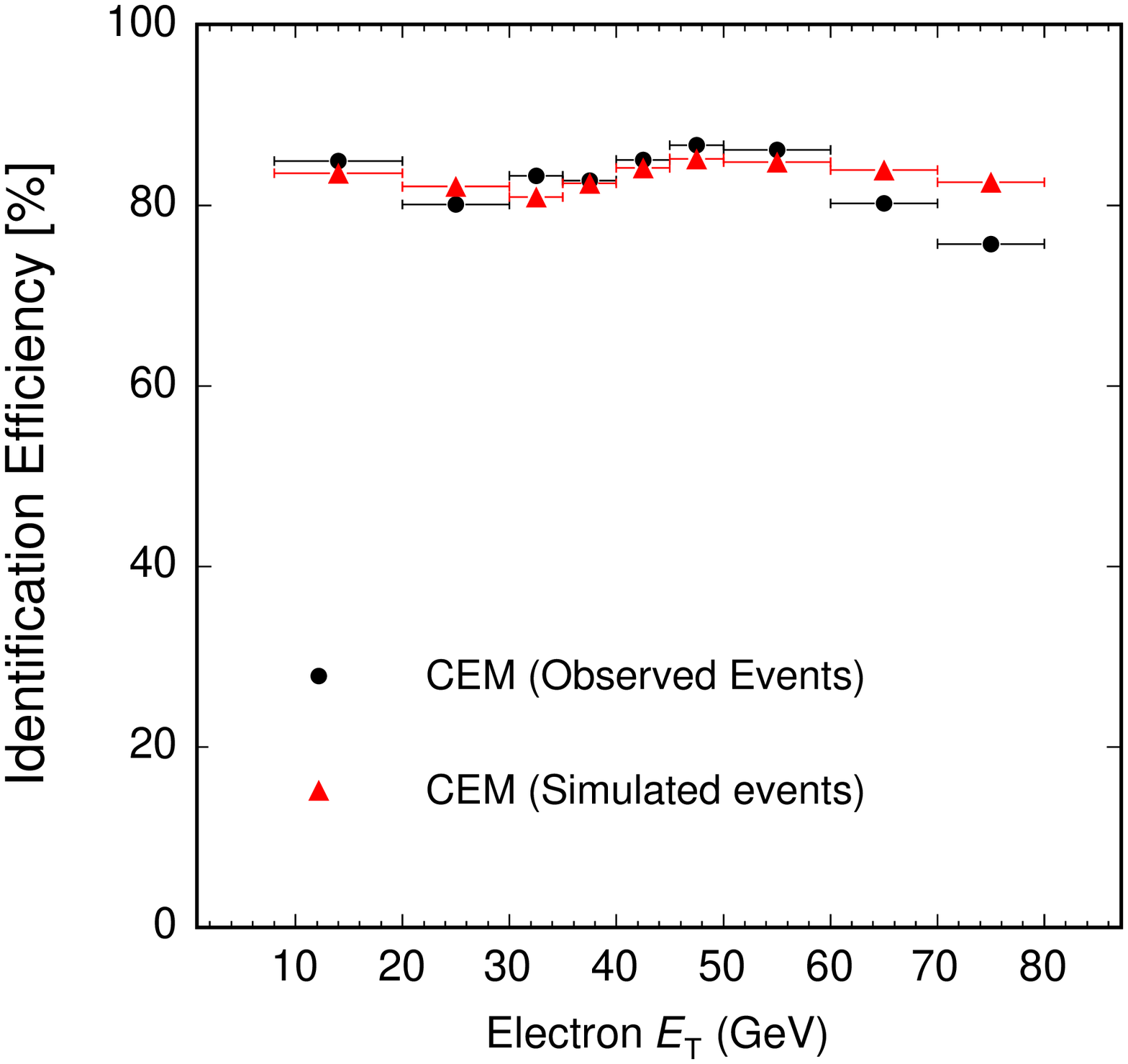}\caption{The identification 
efficiency of tight electrons as a function of $E_{\rm T}$ using the Drell-Yan selection.}\label{fig:electronidet}\includegraphics[width=0.45\textwidth]{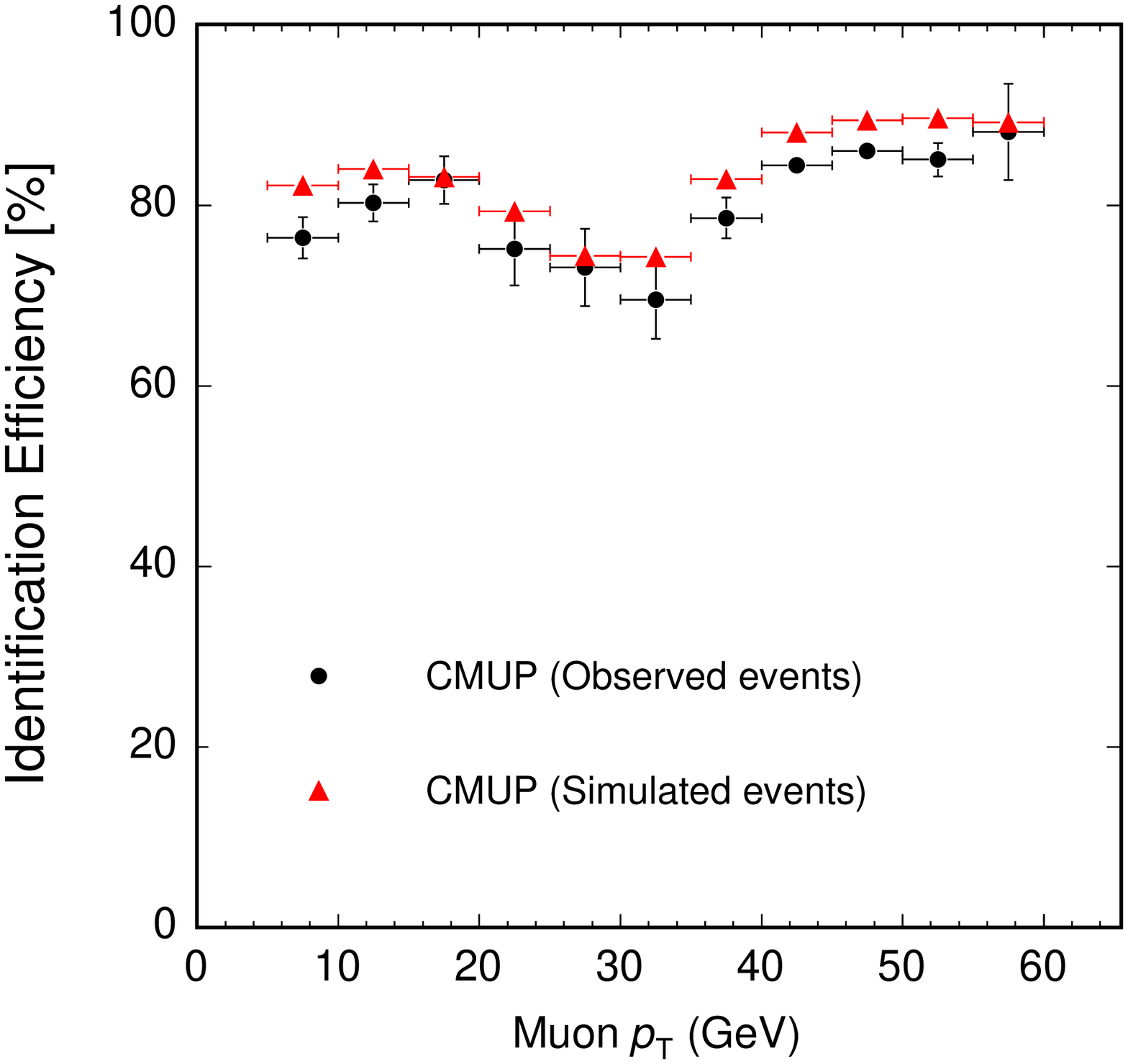}\caption{The identification efficiency (including stub finding efficiency) of CMUP muons as a function of $p_{\rm T}$ using the Drell-Yan selection.}\label{fig:muonidet}\end{center}\end{figure}

For high-\pt\ leptons, we measure the identification efficiencies using
same-flavor, oppositely-charged dilepton candidate events in the invariant mass window from 76 to 106 GeV/$c^2$. We require that at least one of those candidate leptons fulfills all the tight electron or tight CMUP/CMX criteria, defined in Tables~\ref{tab:electrons} and \ref{tab:muons} respectively, and satisfying the trigger requirements. We then measure the efficiencies of our identification criteria on the other candidate lepton. In the case of $Z\rightarrow e^+e^-$ candidates we subtract background using same-charge dilepton events in the mass window. The effect of background subtraction is found to be negligible in the $Z\rightarrow\mu^+\mu^-$ sample~\cite{lepplusjet}.\\ 
\indent The efficiency of low-\pt\ leptons is measured in
Drell-Yan candidate events requiring same-flavor, oppositely-charged leptons with $\Delta\phi (\ell_1
\ell_2)>160^\circ$. 
In order to reject events in which a cosmic ray is reconstructed as a pair of muons, we require the timing of the track hits
in the tracking system to be consistent with particles originating from the center of the detector and moving outwards, and reject
events with significant \met. 
At least one lepton candidate must pass all the identification criteria (to
reduce instrumental and non-prompt background) and must satisfy the 8 GeV/$c$ 
trigger requirements. We then measure the efficiency of the identification variables on
the second lepton candidate in the event. In events in which both leptons pass the trigger requirements, we use both
to determine the efficiency. The remaining background to be subtracted is estimated in events with lepton candidates of the same electric charge. As part of the cross checks and systematic uncertainties evaluation, we also verify the results using $J/\psi$ and $\Upsilon$ candidate events and sideband subtraction, except for the isolation cut, as only the Drell-Yan selection gives well-isolated, prompt leptons with good statistics in the full \pt\ range. The resulting total identification efficiency ranges between 75\% for
 forward electrons and 80\% for central electrons to 90\% for most muon categories.\\
\indent In both observed events and Monte Carlo (MC) simulated events, we check for possible dependence of the efficiency of identifying leptons on additional factors: the number of primary vertices, the geometry of the detector, and changes in the detector performance and/or configuration over time. We include deviations as part of the uncertainty on lepton identification efficiency measurements. 
Fig.~\ref{fig:electronidet} and Fig.~\ref{fig:muonidet}  
show examples of $E_{\rm T}$ (\pt) dependence in observed and simulated events.
The dependence is mainly caused by photons radiated by the leptons; due to the \pt\ spectrum of Drell-Yan events this effect is most visible in the 20-30 GeV range with our selection. The presence of extra photons means the isolation requirement (and for the muons, also the $E_{\rm{EM}}$ requirement) is not fully efficient in that \pt-range for Drell-Yan events. This effect is adequately described by the Drell-Yan simulation. For very high \pt\ electrons the efficiency measured in  observed events is lower than the one measured in Drell-Yan simulated events due to the $E/p$ cut becoming inefficient. 
For CMUP muons the efficiency measured in observed events and in simulated events shows the same dependency with respect to the muon transverse momentum. The efficiency measured in observed
events is lower than the one measured in simulated ones because of mismodelling of multiple scattering.
 Discrepancies at the low-\pt\ end are caused by non-isolated, non-prompt background. 
Because the MC does not completely reproduce the identification efficiency found in the observed events,
 we define a scale factor ($S_{ID}$) as the ratio of the identification efficiency measured in the  observed events to the identification efficiency 
found in the simulated samples. Typical scale factors applied to the MC predictions lie between 0.9 and 1.0 and are \pt\ and $E_{\rm T}$ dependent.

\subsection{Probability of hadrons to be misidentified as leptons}
\label{fake}

\begin{figure}[h]
\begin{center}
\includegraphics[width=0.45\textwidth]{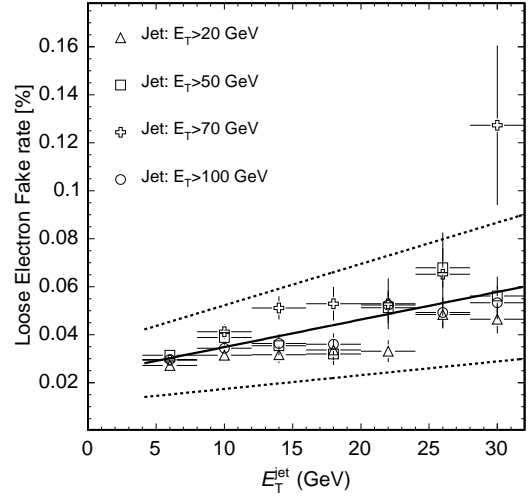}
\caption{Fake rate of loose central electrons. The solid line represents the linear fit to the average and the dotted lines are
$\pm$ 50$\%$.\label{fig:fakeTCE}}
\end{center}
\end{figure}

\begin{figure}[h]
\begin{center}
\includegraphics[width=0.45\textwidth]{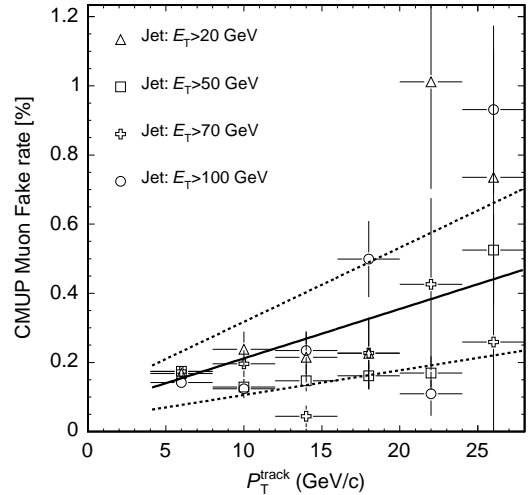}
\caption{Fake rate of CMUP muons. The solid line represents the linear fit to the average and the dotted lines are
$\pm$ 50$\%$. \label{fig:fakeCMUP}}
\end{center}
\end{figure}

A jet 
of hadrons is defined as a cluster of energy in the calorimeter and reconstructed using a 
fixed cone algorithm ($\Delta R$ = 0.4). A jet 
can be misidentified as an electron if it consists of an energetic track
pointing to a large energy deposit in the electromagnetic calorimeter.
Charged kaons and pions with a late shower in the hadronic calorimeter, or those that decay in flight, can also mimic muons. \\
\indent We use reconstructed jets to estimate the probability to misidentify
them as electrons. 
In the study of misidentified muons, we use tracks with $E_{\rm T}^{\rm cone} <$ 4 GeV (called ``isolated tracks''). The isolation is required to reduce the dependence on the sample composition.
In the following we will refer to these jets and tracks as ``fakeable
objects''.
Since such fakeable objects originate from hadrons, we use the four data samples
collected with jet-based triggers to measure their
misidentification probability. We expect only a negligible contribution from inclusive $W$ 
and $Z$ production with the gauge bosons decaying into leptons and do not apply any corrections. To avoid a trigger-induced bias, 
we remove the highest $E_{\rm T}$ jet from the collection of fakeable objects.\\
\indent The misidentification probability, or fake rate, is calculated
as the ratio of the number of identified lepton candidates over the number of fakeable objects. It is
 parametrized as a function of the transverse energy (transverse momentum) of the jet (isolated track)
and averaged over the four jet data samples. The results for one of the electron and muon categories are shown in Fig.~\ref{fig:fakeTCE} and Fig.~\ref{fig:fakeCMUP} respectively.
The  probability for misidentifying hadrons as muons is higher than that for electrons since the muon-type fakeable object 
is based on an isolated track and thus more likely to pass our identification cuts. The application of these rates in the analysis is described in the next section.
An uncertainty of 50\% is assessed from the variation
in the fake rates measured in the different jet data samples.

\section{Samples of observed and simulated events}
\label{dataSample}
\subsection{ Sample of observed events}
The data used in this analysis were collected between March 2002 and February 2006 via electron-based  
and muon-based triggers. The former requires one central ($|\eta|<1$) electron with $E_{\rm T} >$ 18 GeV,
whereas the latter requires one \pt\ $ >18 $ GeV/$c$ central muon with a stub
in both the CMU and CMP or in the CMX chambers.
The data correspond to an integrated luminosity of 1.0 fb$^{-1}$ and 0.7 
fb$^{-1}$ for the samples based on the electron 
and muon triggers, respectively. 
\subsection{Background Samples}
 In the search based on three leptons and
missing transverse energy the SM backgrounds  
are 
 $W\gamma$, $WZ/\gamma^*$, $ZZ/\gamma^*$, 
$t\bar{t}$ and Drell-Yan production, along with hadrons misidentified 
as leptons. The $b\bar{b}$ contamination is suppressed
because the soft and typically non-isolated leptons from $B$ 
decays are rejected by our lepton selection.
 The first set of backgrounds are estimated using a  Monte Carlo technique,
whereas the contribution from misidentified hadrons is measured using observed events (Section~\ref{leptonid}).
\indent The simulated samples are generated using \textsc{pythia}~\cite{pythia}
version 6.216 with the underlying event model tuned to the CDF observed events ~\cite{tunea}. 
In the case of the $WZ$ sample, \textsc{pythia} 
is used only for the parton showering and the hadronization of events that are 
generated with the leading-order matrix element program \textsc{madevent}~\cite{madgraph}. \\
\noindent All simulated background samples were run through the full CDF detector simulation,
 which is based on the
\textsc{geant}~\cite{geant} framework, 
and the same reconstruction algorithm~\cite{cdfsim} that is used for the observed events. 
All simulation-driven background estimates are corrected for the different trigger efficiency (see for instance~\cite{trackeff}) and 
identification efficiency measured in observed events
with respect to the one in simulated events (Section \ref{leptonid}).
An additional correction factor ($S_{\rm conv}$) is needed
for the Drell-Yan production, as explained in the next section.\\
 To avoid overestimation of the background due to hadrons misidentified
as leptons,  we require each identified 
lepton in simulated events to originate from the hard interaction (this does not apply to the $t\bar{t}$ background
where we only ask for three electrons or muons). 

\subsubsection{Drell-Yan}
\label{conversion}
\begin{figure}[h]
\begin{center}
\includegraphics[width=5cm,angle=270,clip=]{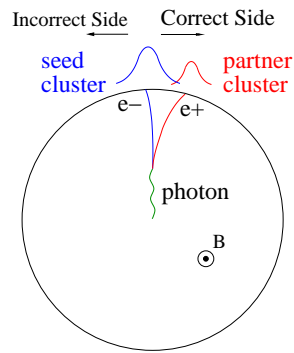}
\caption{Sketch of the r$-\rm{\phi}$ view of a photon conversion signature with CES clusters locations.
The magnetic field $B$ is along the $z$ direction.}
\label{fig:convex}
\end{center}
\end{figure}

\begin{figure}[htb]
\begin{center}
\centerline{\includegraphics[width=.4\textwidth]{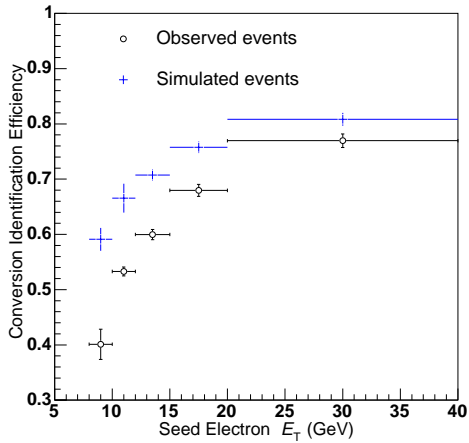}}
\caption{Conversion identification efficiency as a function of the seed electron $E_{\rm T}$.}
\label{fig:conv}
\end{center}
\end{figure}

Events from $Z/\gamma*\rightarrow \ell\ell$ constitute a background to our search if an additional lepton is present
in the event. In this section we present the estimate of this background contribution when the third lepton comes from
a photon radiated from one of the primary leptons, and has converted into an $e^+e^-$ pair.
\indent 
In order to measure the efficiency of the conversion identification 
algorithm described in Section \ref{leptonid} we collect a pure sample of candidate conversions using a calorimeter based approach which
does not rely on tracking information.
The sample consists of identified electrons with {\pt} larger than 8 GeV/$c$ 
(called ``seed electrons'') accompanied by an additional cluster found in 
the shower max detector.
Since photons convert into oppositely-charged electrons \cite{positrons}, 
we can predict the possible $\phi$ location of the cluster based on the charge of the seed electron. 
In Fig.~\ref{fig:convex} 
the ``correct'' and ``incorrect'' sides with respect to the seed electron are defined. 
Furthermore the electrons from $\gamma$ conversions are expected to have the same $z$ coordinate at the CES, since the 
magnetic field $B$ is along the $z$ direction.
Based on this, a candidate photon conversion is a seed electron accompanied by a CES cluster located 
on the ``correct'' side and having $|\Delta z_{\mathrm{seed,cluster}}|< $ 
20 cm.
 In order to improve the purity of the sample of candidate conversions, we reject events in which the seed electron 
comes from a $W$ and is accompanied by a bremsstrahlung photon  
by requiring {\met} to be less than 10 GeV. Furthermore, if the
invariant mass of the seed electron and a second
same-flavor lepton in the event falls in the range from 50 to 106 GeV/$c^2$, the event is considered non-conversion background ($Z$ + bremsstrahlung photon) and rejected. 
Events in which the bremsstrahlung  photon converts are
suppressed by rejecting electrons
having the sum of the measured energy deposit in the electromagnetic calorimeter larger
than the  corresponding
track momentum.
Several other backgrounds mimic the conversion candidate signature,
such as electrons accompanied by a $\pi^{0}$ (decaying into $\gamma\gamma$) or a $K^{\pm}$ (decaying in the detector and producing a shower in the electromagnetic calorimeter as well as in the hadronic calorimeter), or photons from extra interactions
and jets. These components of the background are expected to contribute equally to the ``correct''
and ``incorrect'' sides. Consequently, they can be estimated  by the number of events with clusters
on the ``incorrect'' side. We measure the remaining background in the incorrect side through a fit and  subtract it from the signal.\\ 
\indent The results of the measurement 
performed on the observed and simulated events are shown in Fig.~\ref{fig:conv}.
The sources of inefficiency are mainly track reconstruction inefficiency in the region 
of low \pt, given the asymmetric nature of conversions, and rejections 
due to the thresholds on $D_{xy}$ and $\Delta\cot\theta$. 
Several systematic uncertainties affect the measurement of the conversion 
identification
efficiency, the most significant being the 
uncertainty on the normalization of the background and $\eta$ 
dependence of the efficiency. The total uncertainty is 30$\%$ \cite{alon}. \\
\indent The conversion identification efficiency is lower in observed
than in simulated events. To take this effect into account, we rescale
the contribution of simulated events by $S_{conv}$, the ratio of 
the conversion identification inefficiency in observed events over the inefficiency in
simulated events. We use the inefficiency rather than the efficiency since 
 $S_{conv}(E_{\rm T})$, is applied to electrons
originating from a non-identified conversion in $Z/\gamma^* \rightarrow \ell\ell\ \gamma \rightarrow \ell\ell\ e^{+}e^{-}$. Only electrons
for which the partner track \pt\ is larger than 0.7 GeV/$c$ are corrected.

\subsubsection{Background due to hadrons misidentified as leptons\label{FAKE}}
In order to estimate the background  contribution from events with two leptons and a misidentified lepton~\cite{useFake},
we use the search data sample itself. We select dilepton events with 
at least one additional fakeable object separated from either identified leptons by $\Delta R > 0.4$.
The number of observed events containing two identified leptons and one  fakeable object is then scaled by the probability 
for the fakeable object to be misidentified as a lepton. 
We take into account the fact that  there may be multiple fakeable objects per event. 
\subsection{SUSY samples generation}
The chargino-neutralino cross section depends on the squark mass as can be inferred from Fig.~\ref{fig::pro}, whereas
the branching ratio into three leptons and \met\ depends on the slepton masses. The chargino-neutralino
scenario adopted to guide 
this trilepton analysis is taken from an mSUGRA model (referred to as the 
benchmark point). The benchmark point is characterized by $m_{1/2}$ = 180 GeV/$c^2$,
$m_{0}$ = 100 GeV/$c^2$, $A_{0}$ = 0, $\tan\beta$ = 5, and $\mu >$ 0. The parameters  
$m_{1/2}$ and $m_{0}$ indicate the unified 
gaugino and scalar masses, $A_{0}$ is the unified trilinear coupling of the theory,
 $\tan\beta$ the ratio of the vacuum expectation value of the Higgs doublet coupling to the
up-generation over the one of the Higgs doublet coupling to the
down-generation, and  $\mu$ is the higgsino coupling. 
This benchmark point yields a typical mass spectrum above the LEP chargino mass limit, with charginos of 113 GeV/$c^2$, an LSP of 65 GeV/$c^2$, and 
the lightest stau ($\tilde{\tau_1}$) of 125 GeV/$c^2$. The NLO production
cross section calculated using  \textsc{prospino 2.0}~\cite{prospino} is $\sigma = 0.64\pm0.06$ pb 
and the branching ratio into three leptons is 25$\%$, 
as obtained from \textsc{pythia}. 
 The SUSY simulation sample 
is generated with \textsc{pythia} version 6.216.
In this mode, \textsc{pythia} obtains 
 the masses at the
 electroweak scale from the routine \textsc{isasugra} (\textsc{isajet}~\cite{isajet} version
 7.51). \\
While the benchmark point is used to study the event kinematics of chargino-neutralino associated production,
three additional scenarios are used to fix squark and slepton masses and to interpret the results of  our search.
The modeling of the non-mSUGRA models is done by using \textsc{softsusy} 2.0.7~\cite{softsusy} as the input to \textsc{pythia} 6.325, 
using the SUSY Les Houches Accord~\cite{slha} framework.    
The SUSY contribution
is corrected to take into account the different identification efficiency 
measured in observed and simulated events,
the same way as for the backgrounds.

\section{Event Selection}
\label{anaCuts}
The samples of observed and simulated events are divided into four non-exclusive channels:  $ee\ell$, $e\mu\ell$, $\mu\mu\ell$, and $\mu e\ell$, in which
the first lepton listed is the one which passed the trigger requirements, and $\ell$ is an electron or muon. The lepton selection accepts also $\tau$ leptons, when they decay to electrons or muons.\\
\indent In the  $ee\ell$ and $e\mu\ell$ subsets, each event
must contain at least one tight central electron with $E_{\rm T} > $ 20 GeV, consistent
with the trigger object. The second lepton listed is either a loose electron or a plug electron 
with $E_{\rm T}$ $ > $ 8 GeV, or a muon with \pt\ $ > $ 8 GeV/$c$ (10 
GeV/$c$
for CMIO's). In the  $\mu\mu\ell$ and 
$\mu e\ell$ subsets \cite{anadi}, at least one lepton
must be a   CMUP or CMX muon with \pt\ $ > $ 20 GeV/$c$, and the second
lepton can be either a loose central or plug electron with $E_{\rm T} > $ 8 GeV, or a muon with \pt\ $ > $ 5 GeV/$c$ (10 GeV/$c$ for CMIO's). The third lepton listed can
be from any of the above categories with a common \pt\
($E_{\rm T}$) threshold of 5 GeV/$c$ (5 GeV), except for CMIO's for which we always
require \pt\ $ > $ 10 GeV/$c$. 
\subsection{Preselection}
Based on the expected topology of chargino-neutralino events, 
we  
require all leptons to originate from the primary vertex, $|\Delta z (\ell_i,\mbox{primary vertex})|<$ 4 cm and $|\Delta z (\ell_i,\ell_j )|<$ 4 cm, 
and to be separate in $\eta$-$\phi$ space with $\Delta R >$ 0.4.\\
\indent
The energy of candidate jets with $E_{\rm T} >$ 5 GeV and 
within $|\eta|<2.5$ is corrected to take into account
the geometry of the calorimeters and the nonlinearity of their 
response~\cite{florencia}. 
We do not include candidate jets that have a high electromagnetic to total energy ratio, consistent with being electrons. Each candidate jet is required to be far from all identified leptons in the event ($\Delta R > $ 0.4). In the particular case of the $e\mu\ell$ channel,
 we reject events in which either the second or the third
lepton is within 20 degrees from the jet axis. \\
\indent The missing transverse energy, reconstructed from
calorimeter towers with transverse energy larger than 0.1 GeV within
$|\eta|<3.6$, is corrected for muons because muons 
leave only small deposits of energy in the calorimeter. In the \met\  calculation, we take this 
effect into account by
subtracting the transverse momenta of identified muon tracks
from the \met, after adding the average muon energy measured in 
the calorimeter and projected into the transverse plane.
We suppress events with mis-measured \met\ by requiring the \met\ to
be separated by at least 2 degrees in azimuth 
from the vectorial sum of the transverse momenta of the two highest-\pt\ leptons,
in events in which the second lepton is a muon. 
This selection is designed to 
reject potentially problematic Drell-Yan events where the leptons 
and the lepton energy is  mis-measured, producing
 missing transverse energy along its direction.
The underestimation or overestimation of the energy of a jet causes a spurious energy
imbalance in the event which affects the value of missing 
transverse energy.
In order to remove such events, we require the smallest 
angle between \met\ and the axis of any candidate jet to be $\Delta \phi 
> 20$ degrees in the $\mu\mu\ell$ and $e\mu\ell$ channels. \\
\indent In observed events with muons, cosmic rays are 
identified (and rejected) as two tracks aligned in the
transverse plane satisfying quality and matching requirements. 
To reduce further the cosmic background in the $\mu\mu\ell$
and $\mu e\ell$ channels, 
we veto events in which the two 
highest \pt\
muons exhibit a three-dimensional angular separation larger than 178 degrees.\\

\subsection{Kinematic selection}
\begin{figure}
\begin{center}
\includegraphics[width=0.4\textwidth]{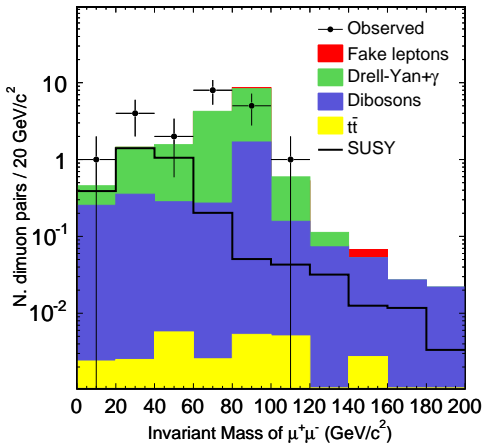}
\caption{Invariant mass of same flavor, oppositely charged leptons in $\mu\mu\ell$ events. The SM backgrounds are stacked while the
benchmark SUSY signal is superimposed. Observed events are shown as points with error bars indicating the statistical uncertainty.}
\label{fig:CRtri}\end{center}\end{figure}

In order to achieve the best sensitivity, several event
selection criteria are applied to reject the backgrounds.\\
\indent An important discriminating variable is the invariant mass of same-flavor, oppositely-charged leptons.
The on-shell component of the $Z$ production is suppressed 
by rejecting 
events with two leptons of the same flavor with a combined invariant mass 
in the window  of 76 to 106 $\mbox{GeV}/c^2$. This selection also 
reduces the 
otherwise indistinguishable $WZ$ background.
Similarly, the low mass resonances such as $J/\psi$ and $\Upsilon$
are removed by requiring a dilepton invariant mass larger than
15 $\mbox{GeV}/c^2$. The latter value is raised to 20 $\mbox{GeV}/c^2$ for
$ee\ell$ and $e\mu\ell$ events. Fig.~\ref{fig:CRtri} shows the invariant 
mass of muons pairs in trilepton
events. \\
\indent In chargino-neutralino events in which the supersymmetric particles decay
into leptons, we expect jet activity to come only from initial state
radiation. On the other hand, $t\bar{t}$ events always contain jets, a feature
 which distinguishes them from chargino-neutralino signal events. The  $t\bar{t}$
background is reduced by rejecting events with more than one jet with  
$E_{\rm T} > 20$ GeV.\\
\indent Finally,  SUSY events are characterized by significant
missing transverse energy from the LSP's and the neutrinos. This pattern
differs from Drell-Yan production of charged leptons, where only the $Z\rightarrow \tau\tau$ background exhibits real missing transverse energy. 
 We require \met$ >$ 15 GeV in order to remove the Drell-Yan events outside the $Z$ mass window.\\
\indent The resulting predictions of the signal yields in the benchmark point ($S$)
 and the
accompanying SM backgrounds ($B$) after all cuts are given in Table~\ref{tab:all}.
Typically the dominant background is due to real trilepton processes, while
the $t\bar{t}$ contribution is negligible. 
The $\mu\mu\ell$ channel displays the highest sensitivity defined as $S/\sqrt{B}$. \\
It can be inferred from the table that the prediction of Drell-Yan + $\gamma$ is limited by the available number of simulated Drell-Yan events. This is due to a very high rejection factor for this background, and the resulting large uncertainty is taken into account in the limit calculation.
\begin{table}[h]
\begin{center}
\caption{Expected background and predicted signal for the mSUGRA benchmark point for all channels. The uncertainties shown are statistical only. }\label{tab:all}
\begin{tabular}{ccccc} \hline \hline
& $\mu\mu \ell$    &  $\mu e \ell$                  &   $ee\ell$           & $e\mu \ell$ \\ \hline
Diboson            &  0.20$\pm$0.02         &   0.44$\pm$0.04      &  0.29$\pm$0.02  &  0.17$\pm$0.02\\
Drell-Yan +$\gamma$        &  0.22$\pm$0.11         &   0.14$\pm$0.06      &  0.14$\pm$0.02  &  0.04$\pm$0.04\\
Fake leptons        &  0.20$\pm$0.02         &   0.17$\pm$0.02      &  0.11$\pm$0.01  &  0.05$\pm$0.01\\
$t\bar t$          &  0.01$\pm$0.01         &   0.02$\pm$0.01      &  0.02$\pm$0.01  &  0.02$\pm$0.01\\ \hline
   Total SM.  &  0.64$\pm$0.11         &   0.78$\pm$0.08      &  0.56$\pm$0.03  &  0.28$\pm$0.05 \\ \hline 
Signal             &  1.60$\pm$0.11         &   1.03$\pm$0.06      &  1.22$\pm$0.08  &  0.84$\pm$0.07\\ \hline \hline
\end{tabular}
\end{center}
\end{table}

\section{Systematic Uncertainties}
\label{sys}
There are several systematic uncertainties which
affect the numbers of predicted events and, consequently, the interpretation of the search result. 
The relative contributions vary from channel to channel and from signal to background. In case of the background estimate,
the largest uncertainty originates from the 
  statistical uncertainty on the number of
predicted events.  These uncertainties are not negligible (up to 29\%)
due to the 
 finite sample sizes and are 
included as independent sources of systematic uncertainty in the
limit calculation described in Section~\ref{model}. The  uncertainty from the
background due to  misidentified leptons is determined from the precision of the fake rate measurement and it can be as large as 16\%. 
For most lepton categories the systematic uncertainty due to the scale factors of the identification
efficiency  is a few percent,  except for low-$E_{\rm{T}}$ plug electrons which have a 14\%\ uncertainty.
  The jet energy scale~\cite{florencia} is varied within
its uncertainty to estimate the impact on the jet multiplicity
and on the correction of the missing transverse energy. The effect is between 2\% and 7\% depending on the channel. 
 The integrated luminosity is measured with an accuracy of 6\%~\cite{lumi} and it is used to normalize the  contributions from simulated events.  
The initial state radiation 
(ISR) and final state radiation (FSR) are modeled in the
simulated samples and are subject to the uncertainty of the parton
shower model. The effects of these uncertainties are determined from samples simulated 
with different ISR/FSR content~\cite{topmasspaper}, resulting in variations of up to 4\%\ in selection acceptance. The cross sections and the event kinematics
depend on the momenta of the incoming partons, whose spectra
are parametrized by parton distribution functions (PDF's) obtained from a fit to the data from a number of experiments. 
We calculate the uncertainties on 
background rates by adding in quadrature the differences between each of the 40 CTEQ6~\cite{cteq6m} systematic-variation
eigenvectors and the nominal predictions. The effects on the cross
sections and the acceptances are included.  The resulting uncertainty on
the background rates is 2\%.\\
\indent We also estimate the effect of the uncertainty 
in the theoretical cross section predictions for diboson production (7\%)~\cite{diboson}, and $t\bar t $ production (10\%)~\cite{ttbar}. \\
The individual contributions for background and SUSY signal are summarized in Table~\ref{tab:sys}, and the total systematic uncertainties for each channel are listed in Table~\ref{tab:systot}. 
\begin{table}[h]
\begin{center}
\caption{Summary of systematic uncertainties.}
\begin{tabular}{crr} \hline  \hline  
  & \multicolumn{2}{c} {Resulting variation in}  \\
Source & Signal & Background  \\ \hline
Monte Carlo statistics & 6-10\% & 12-29\%\\ \hline
Hadron misidentification efficiency & - & 9-16\%\\
 Lepton identification efficiency & 2-7\% & 2-14\%\\
 Jet energy scale & 0.3-3\% & 2-7\%\\
Luminosity & 6\% & 4-5\%\\
ISR/FSR & 2-12 \% & 3-4\%\\
 PDF's & 1\% & 2-3\%\\
 Theoretical $\sigma$ uncertainty & 10\% & 4-8\%\\ \hline \hline  
\end{tabular}
\label{tab:sys}
\end{center}
\end{table}

\begin{table}[h]
\begin{center}
\caption{Combination of all systematic uncertainties in Table~\ref{tab:sys} for signal and background for each channel.}
\begin{tabular}{ccccc} \hline \hline  
Systematic uncertainties & $\mu\mu \ell$&  $\mu e \ell$  &  $ee\ell$  &  $e\mu \ell$ \\ \hline
Background & 28\% & 22\%  & 20\% & 31\%\\ 
Signal & 13\% & 16\%  & 14\%  & 14\% \\ \hline   \hline  
\end{tabular}
\label{tab:systot}
\end{center}
\end{table}

\section{Control Samples}
\label{cr}
\begin{figure}
\begin{center}
\includegraphics[width=0.4\textwidth]{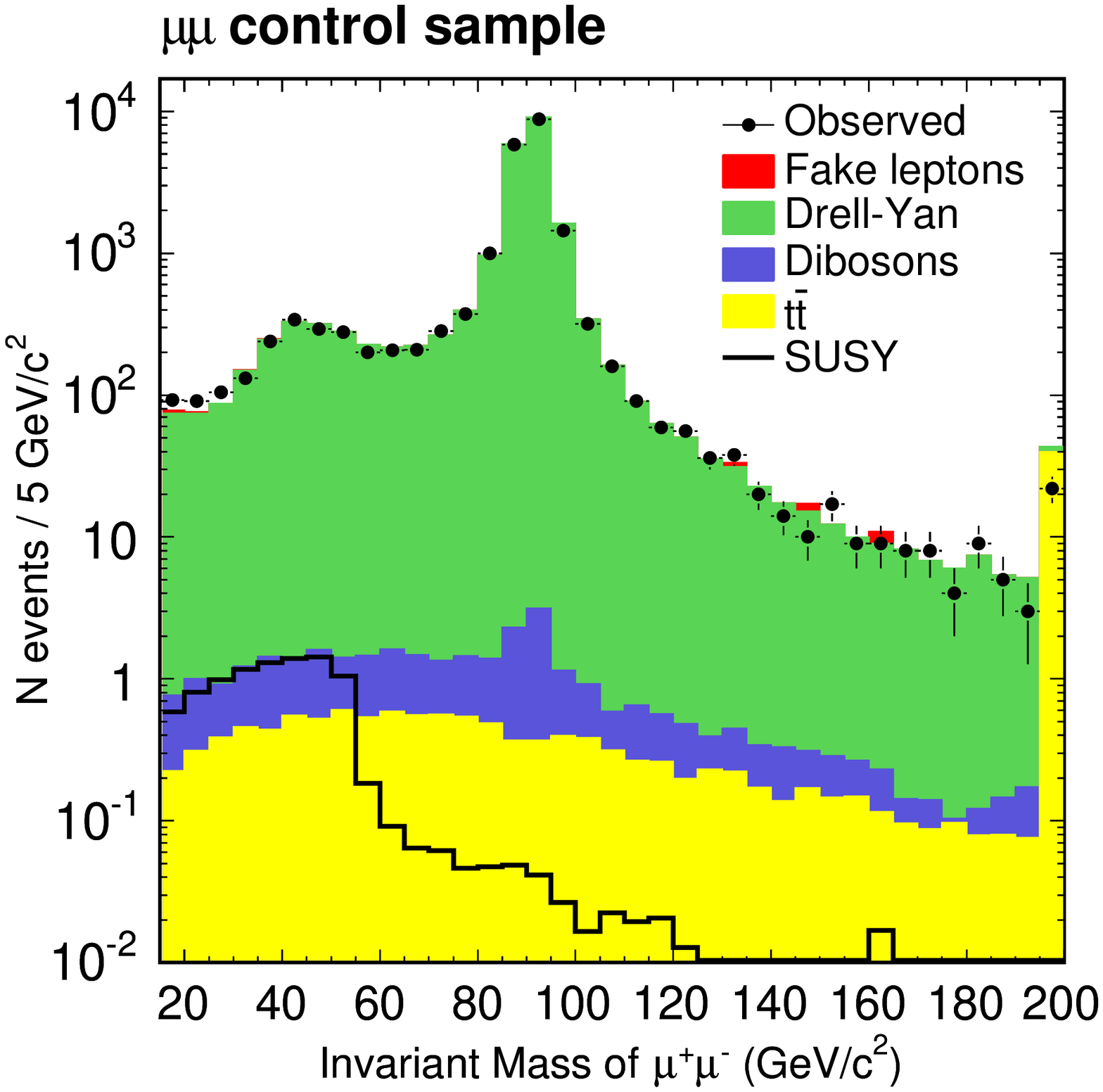}
\includegraphics[width=0.4\textwidth]{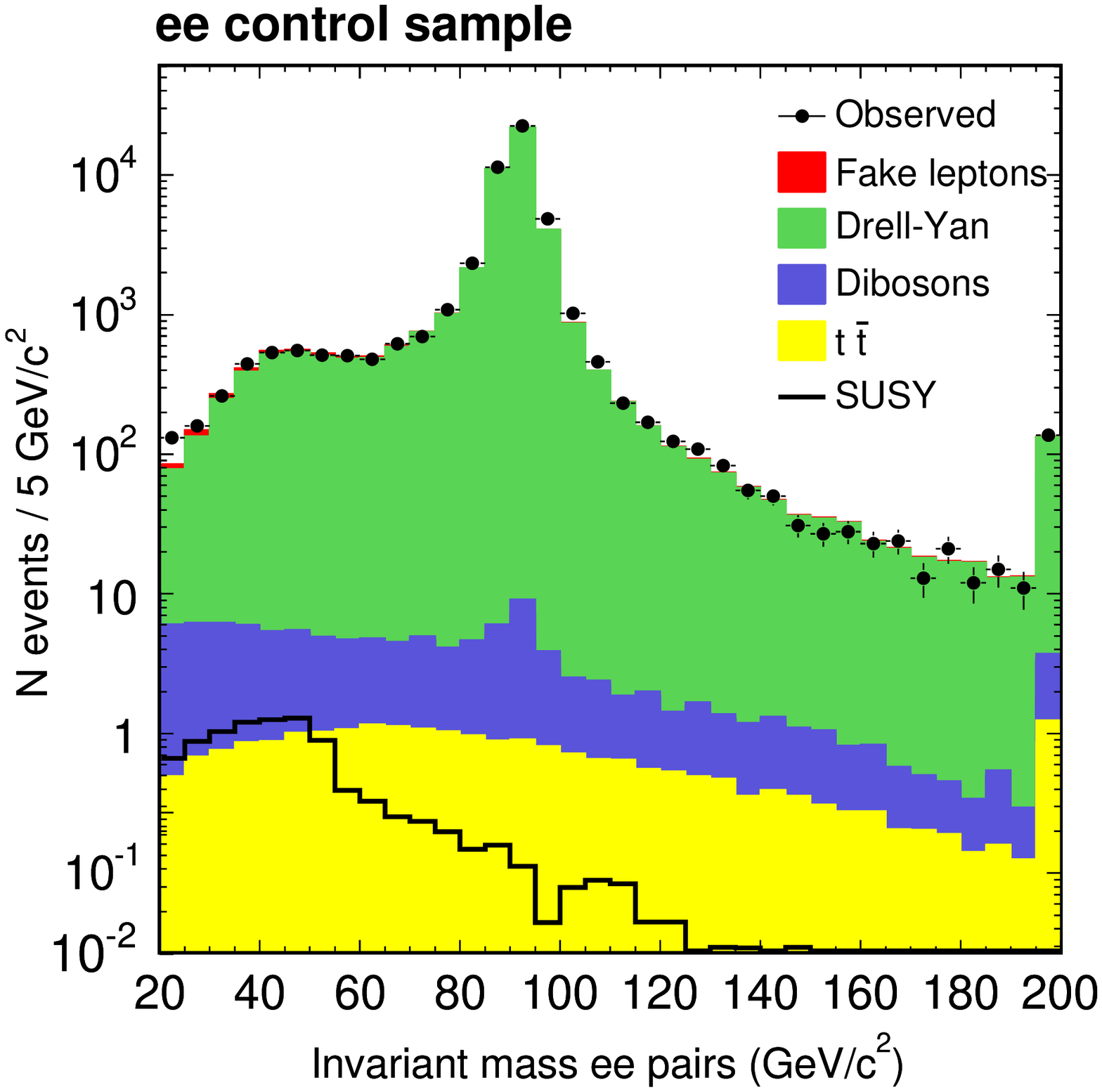}
\caption{Invariant mass of same-flavor lepton pairs. The SM backgrounds are stacked while the
benchmark SUSY signal is superimposed. Observed events are shown as points with error bars indicating the statistical uncertainty (overflows are added to the last bin).}
\label{fig:mass}\end{center}\end{figure}

\begin{figure}
\begin{center}
\includegraphics[width=0.4\textwidth]{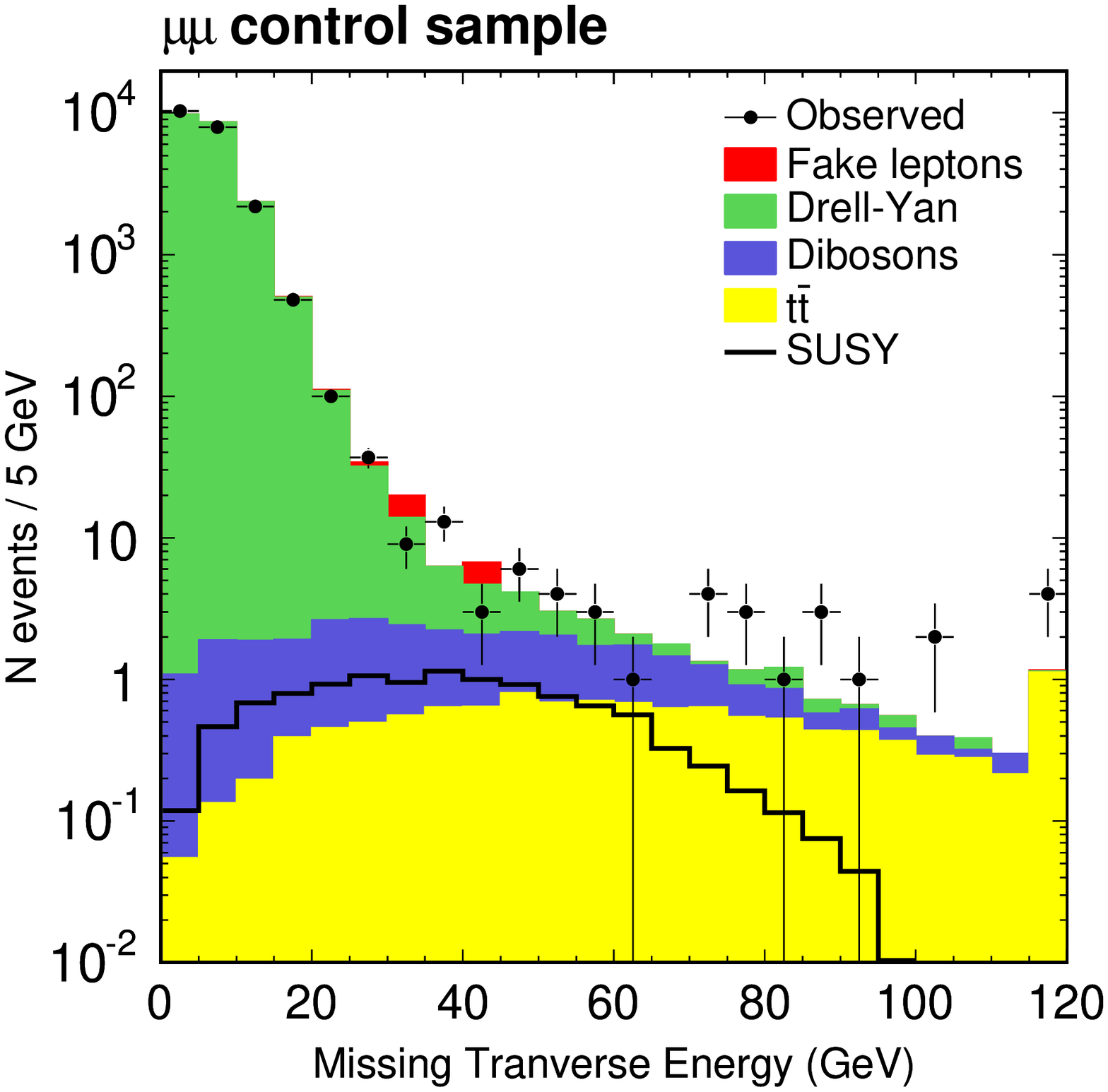}
\includegraphics[width=0.4\textwidth]{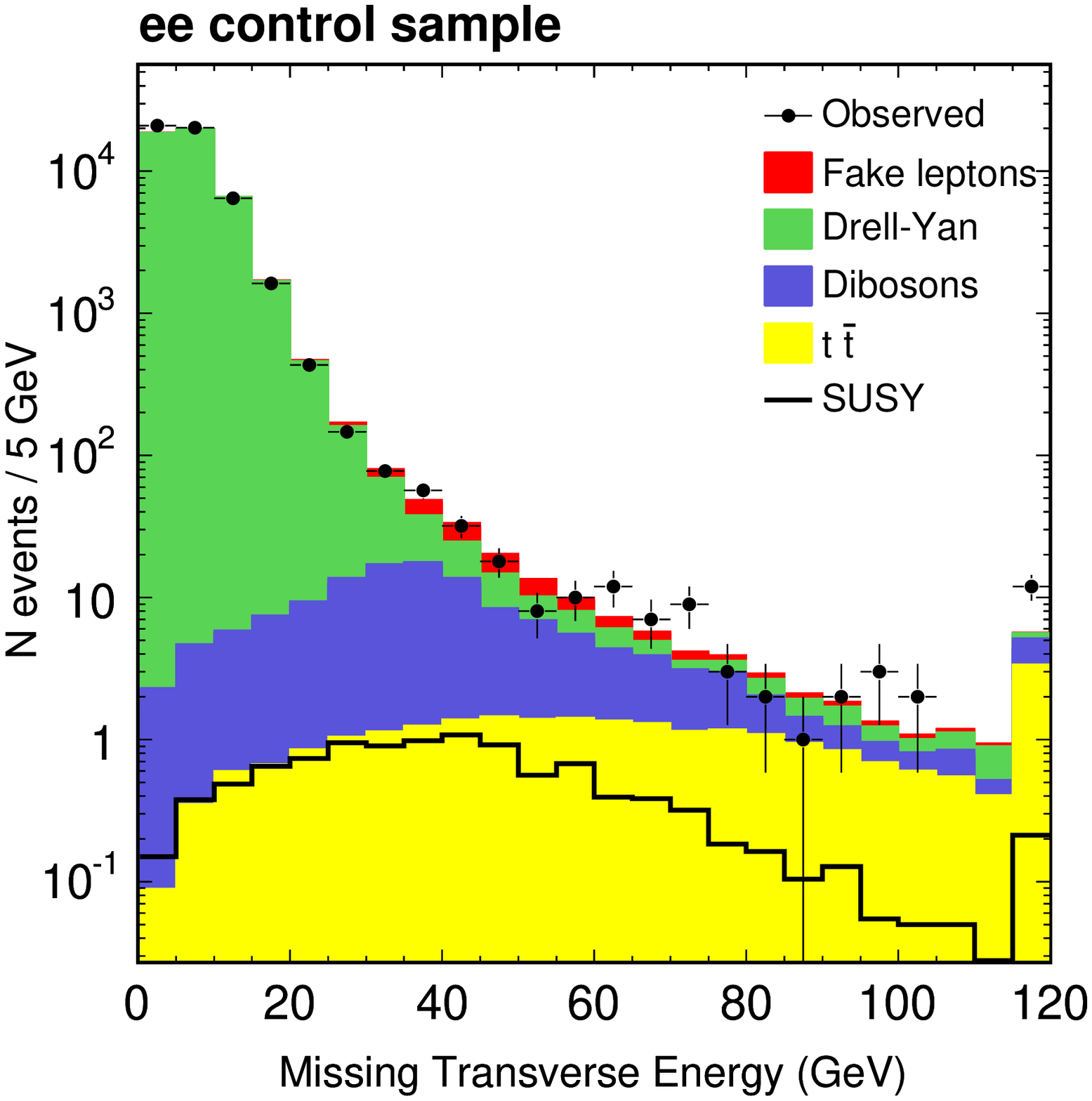}
\caption{Missing transverse energy in $\mu\mu$ and $ee$ events. The SM backgrounds are stacked while the
benchmark SUSY signal is superimposed. Observed events are shown as points with error bars indicating the statistical uncertainty (overflows are added to the last bin).}
\label{fig:metmu}\end{center}\end{figure}

\begin{figure} 
\begin{center}
\includegraphics[width=0.4\textwidth]{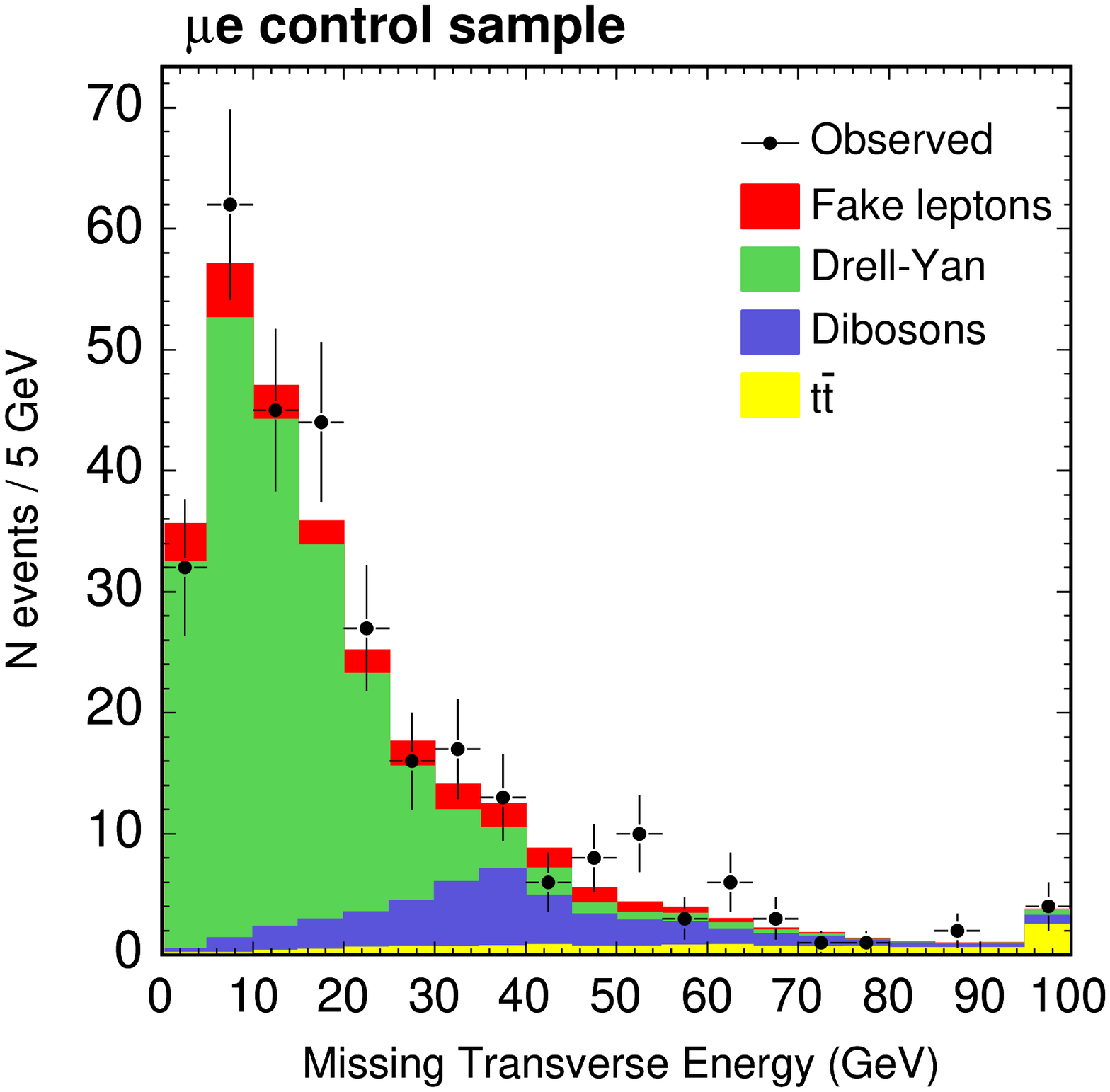}
\includegraphics[width=0.4\textwidth]{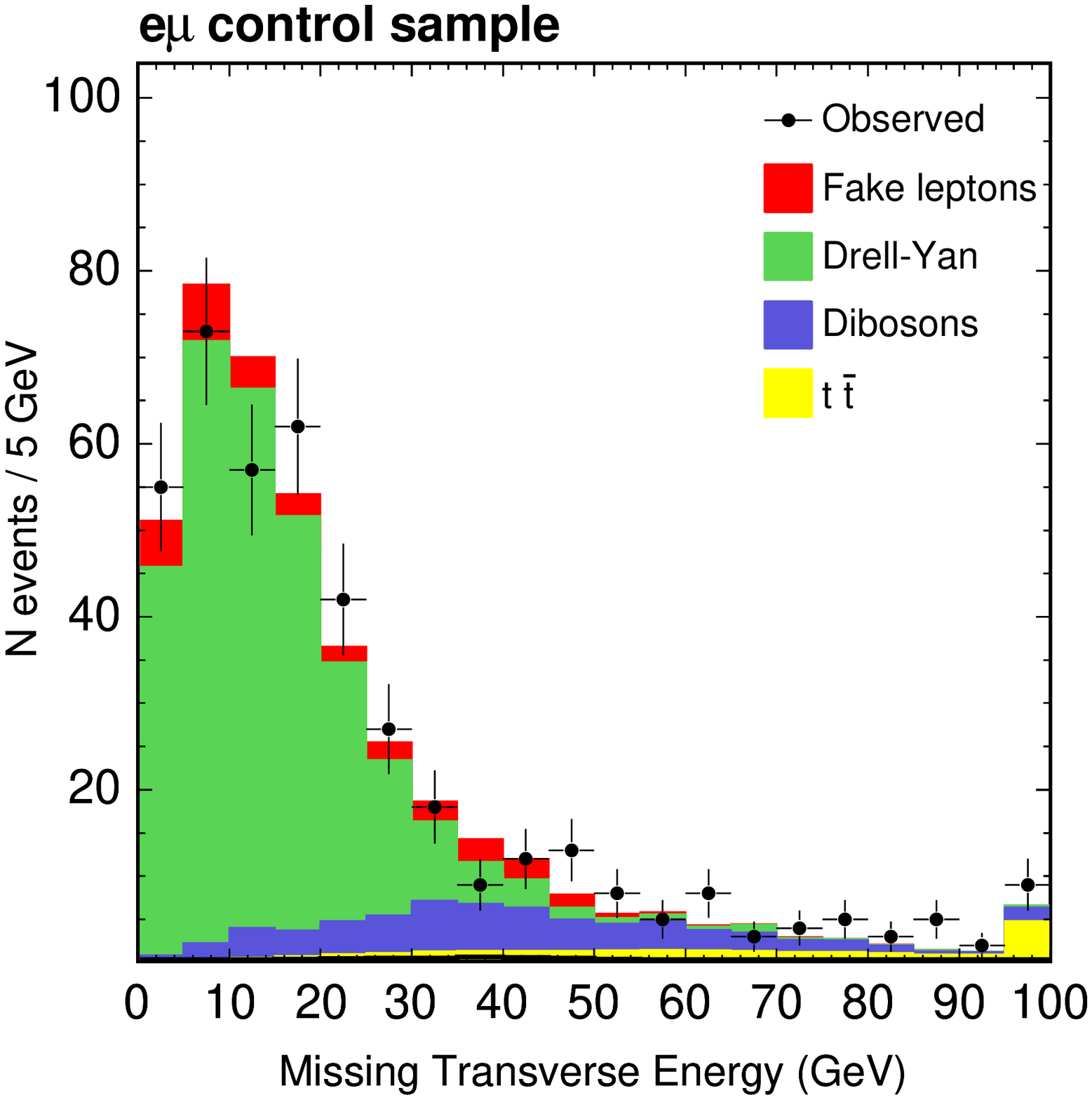}
\caption{Missing transverse energy in $\mu e$ and $e\mu$ events.
The SM backgrounds are stacked while the the benchmark SUSY signal superimposed.  Observed events are shown as points with error bars indicating the statistical uncertainty (overflows are added to the last bin).
The expected benchmark SUSY signal in the $\mu$e and e$\mu$ channels is $\sim$ 3 and $\sim$ 6 events, respectively.
}
\label{fig:mete}\end{center}\end{figure}

\begin{figure} 
\begin{center}
\includegraphics[width=0.4\textwidth]{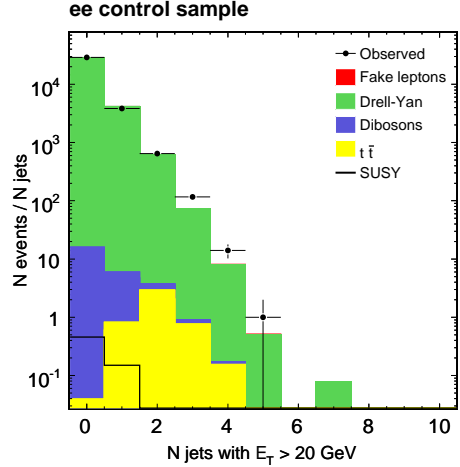}
\caption{Number of jets in $ee$ events with
invariant mass between 76 and 106 GeV/$c^2$. The SM backgrounds are stacked while the
benchmark SUSY signal is superimposed. Observed events are shown as points with error bars indicating the statistical uncertainty.\label{fig:CRZ}}
\end{center}
\end{figure}

We test the SM predictions against the observed events by defining control samples 
in which we expect negligible contributions from SUSY events predicted by 
the benchmark point. We classify each event according to the missing transverse energy, the number of jets, the number of leptons, and, for $ee$ and $\mu\mu$ events, the invariant mass of same-flavor, oppositely-charged leptons. In particular, the subsample of events with 
two leptons is referred to as the ``dilepton control region'', and 
the subsample of events which contain three leptons is referred to as the 
``trilepton control region''. \\
\indent The normalization of the inclusive mass spectra for $ee$ and $\mu\mu$ events, presented in
Fig.~\ref{fig:mass} with the benchmark SUSY signal superimposed, demonstrates good understanding of the
trigger and identification efficiencies along with the measurement of the integrated luminosity.  The quality of the track and jet
reconstruction can be assessed by comparing the missing transverse energy distributions in the observed and in the simulated events,
as illustrated in Fig.~\ref{fig:metmu} and Fig.~\ref{fig:mete}. 
Same-flavor dilepton events are mainly DY~$\mu^+\mu^-$ and DY~$e^+e^-$ as indicated by
the softer \met\ spectrum. In the $\mu e$ and $e\mu $ channels, the broader \met\ spectrum originates from the
leptonic decay of DY~$\tau^+\tau^-$. Only for \met $>$ 40 GeV do other processes become important.
The good agreement between the observed and simulated events shows that the \met\ resolution is simulated well.
The jet multiplicity in  DY~$e^+e^-$ candidate events is compared
to the  predictions based on initial-state radiation
and extra interactions in Fig.~\ref{fig:CRZ}. The event generator \textsc{pythia} reproduces the observed data spectrum well in the region of our interest at low jet multiplicity, whereas for large jet multiplicity 
a NLO simulator with proper parton shower would be needed. 
\indent In  Table \ref{tab:CR} we present examples of 
the numerical comparison between the  observed events and the
total
expected background in 12 of the control samples we investigated: 
\begin{description}
  \item (I) dielectron events
with invariant mass outside the $Z$ window and \met\ $\leq$ 10 GeV,
  \item (II) dielectron events with invariant mass in the $Z$ window and
\met\ $\geq$ 15 GeV,
\item (III) dimuon events in the $Z$ mass window and \met\ $\leq$ 10 GeV,
\item (IV) dimuon events in the $Z$ mass window with at least two jets and \met\ $\geq$ 15 GeV,
\item(V) $\mu e$ events with at least two jets and \met\ $\geq$ 15 GeV,
\item (VI) $\mu e$ events with \met\ $\geq$ 15 GeV,
\item (VII) $e\mu$ events with \met\ $\geq$  15 GeV and $\Delta\phi_{e, \mu}\leq$170$^o$,
\item (VIII) $e\mu$ events with \met\ $\leq$ 10 GeV,
\item (IX) $\mu\mu\ell$ events with \met\ $\leq$ 10 GeV,
\item (X) $\mu\mu\ell$ events in the $Z$ mass window with \met\ $\geq$ 15 GeV,
\item (XI) $ee\ell$ events with \met\ $\leq$ 10 GeV,
\item (XII) $ee\ell$ events in the $Z$ mass window with \met\ $\geq$ 15 GeV and at least two jets.
\end{description}

The trilepton control samples are particularly useful to verify the background from diboson production and misidentified hadrons.
No significant discrepancies are seen between the predictions and the observations.
\begin{table}[h]
\begin{center}
\caption{Examples of control samples as listed in Section~\ref{cr}.
The error on the number of events expected from SM backgrounds includes statistical and systematic
uncertainties. \label{tab:CR}}
\begin{small}
\begin{tabular}{cccccr} \hline \hline
        & \footnotesize Drell-Yan    & \footnotesize Diboson,        & \footnotesize Misid.    & \footnotesize Tot.    & \footnotesize  Observed \\ 
  & \footnotesize     & \footnotesize $t\bar{t}$         & \footnotesize hadrons    & \footnotesize  background   & \footnotesize  \\ \hline
 \multicolumn{2}{c} {\bf\footnotesize 2 leptons:} &&&&\\ 
I       &  2359  &   1.9   &  33        &   $2394\pm314$   & 2422\\ 
II      &  656   &   9.2   &  3.2       &   $669\pm159$    & 638 \\ 
III     &   15587  &   1.5   &  $<$4.5   &   $15588\pm2044$     & 15366\\ 
IV      &   29     &   2.0   &  $<$4.5   &   $31\pm4$        & 31 \\ 
V       &  1.6  &    9.1   &  1.0       & 11.7  $\pm$   2.1   & 7 \\ 
VI      &  76   &   49   &  12.5      & 138  $\pm$  22  & 151 \\ 
VII     &  22   &    16   &  1.2     & 38  $\pm$ 6     & 44 \\  
VIII    &  67  &     1.9  &  5.7     & 74  $\pm$ 9     & 62 \\  
\hline
 \multicolumn{2}{c} {\bf\footnotesize 3 leptons:} &&&&\\ 
IX     &  3.9  &   0.1  &  0.3       & 4.3  $\pm$  1.3   & 4 \\ 
X      &  0.5  &   1.1   &  0.5       & 2.1  $\pm$  0.5   & 2 \\ 
XI     &  3.3  &   0.2   &  0.4       & 3.9  $\pm$  0.6   & 4 \\  
XII    &  0.01 &   0.01  &  0.04      & 0.06  $\pm$ 0.02  & 0 \\  
\hline \hline
\end{tabular}
\end{small}
\end{center}
\end{table}

\section{Results}
\label{results}
\begin{figure}
\begin{center}
\includegraphics[width=0.4\textwidth]{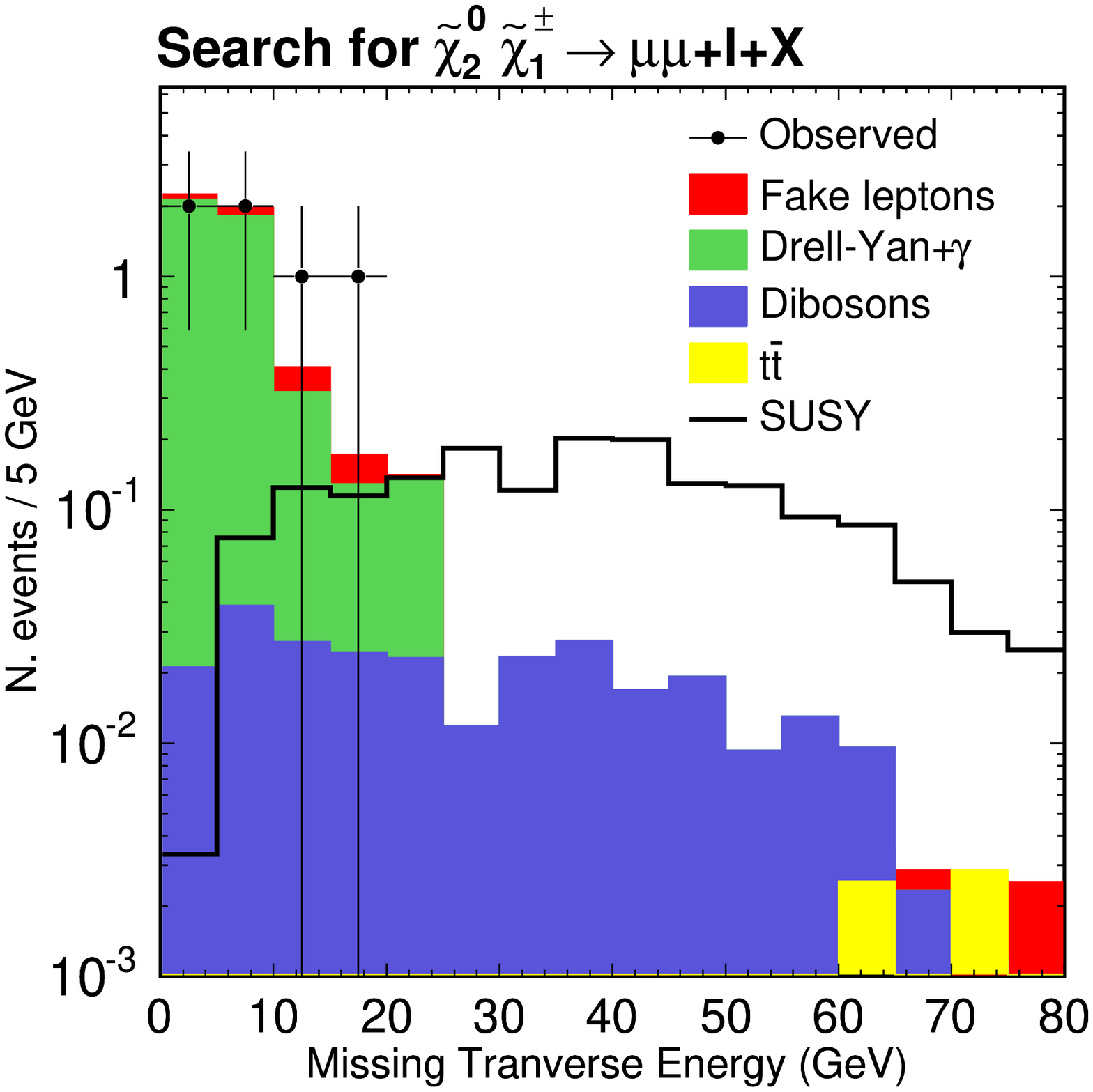}
\includegraphics[width=0.4\textwidth]{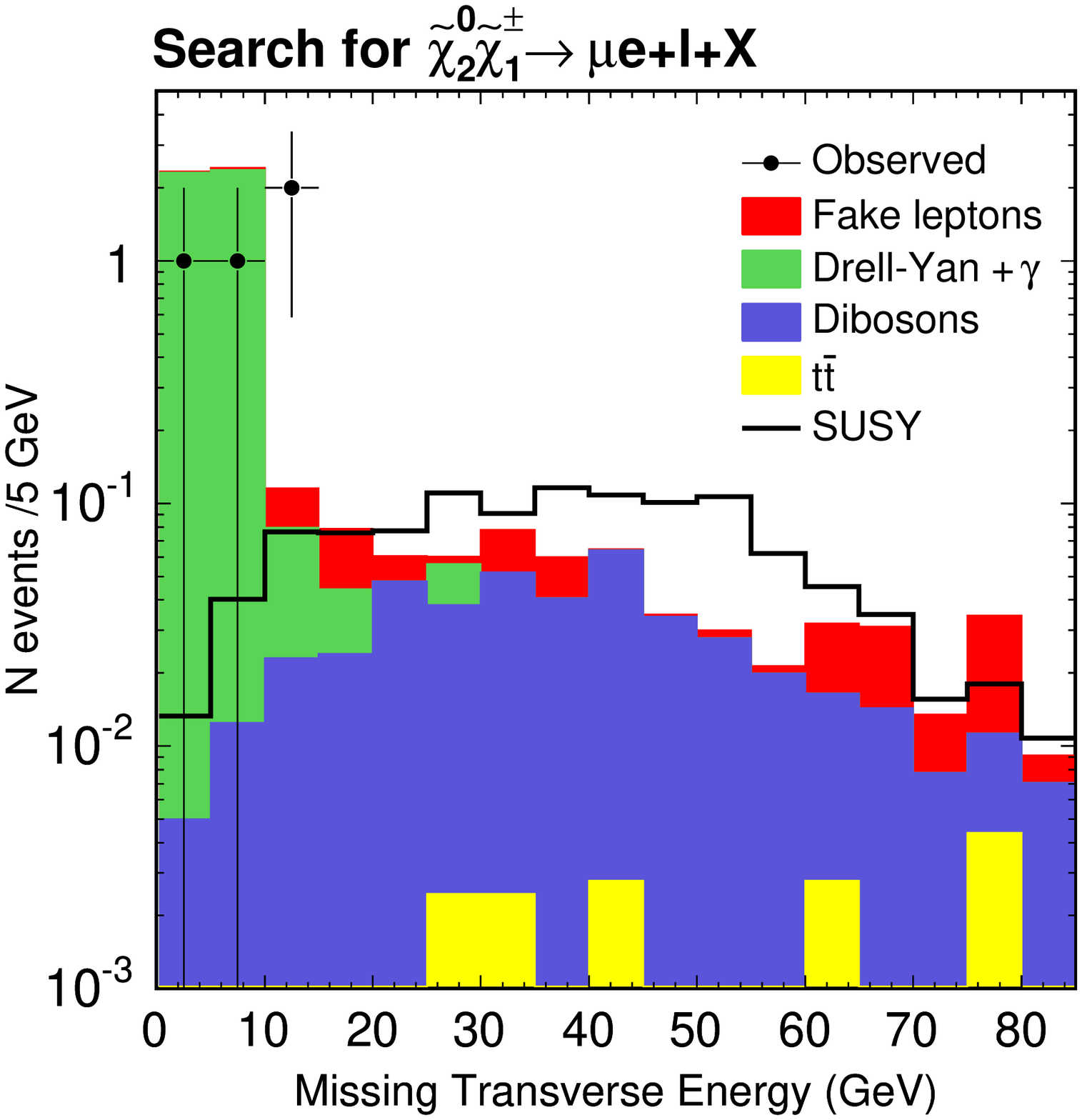} 
\caption{Missing transverse energy before the final cut of 15 GeV in $\mu\mu\ell$ and $\mu e\ell$. The SM backgrounds are stacked while the benchmark 
SUSY signal is superimposed. Observed events are shown as points with error bars indicating the statistical uncertainty .}
\label{fig:muons}\end{center}\end{figure}

\begin{figure}
\begin{center}
\includegraphics[width=0.4\textwidth]{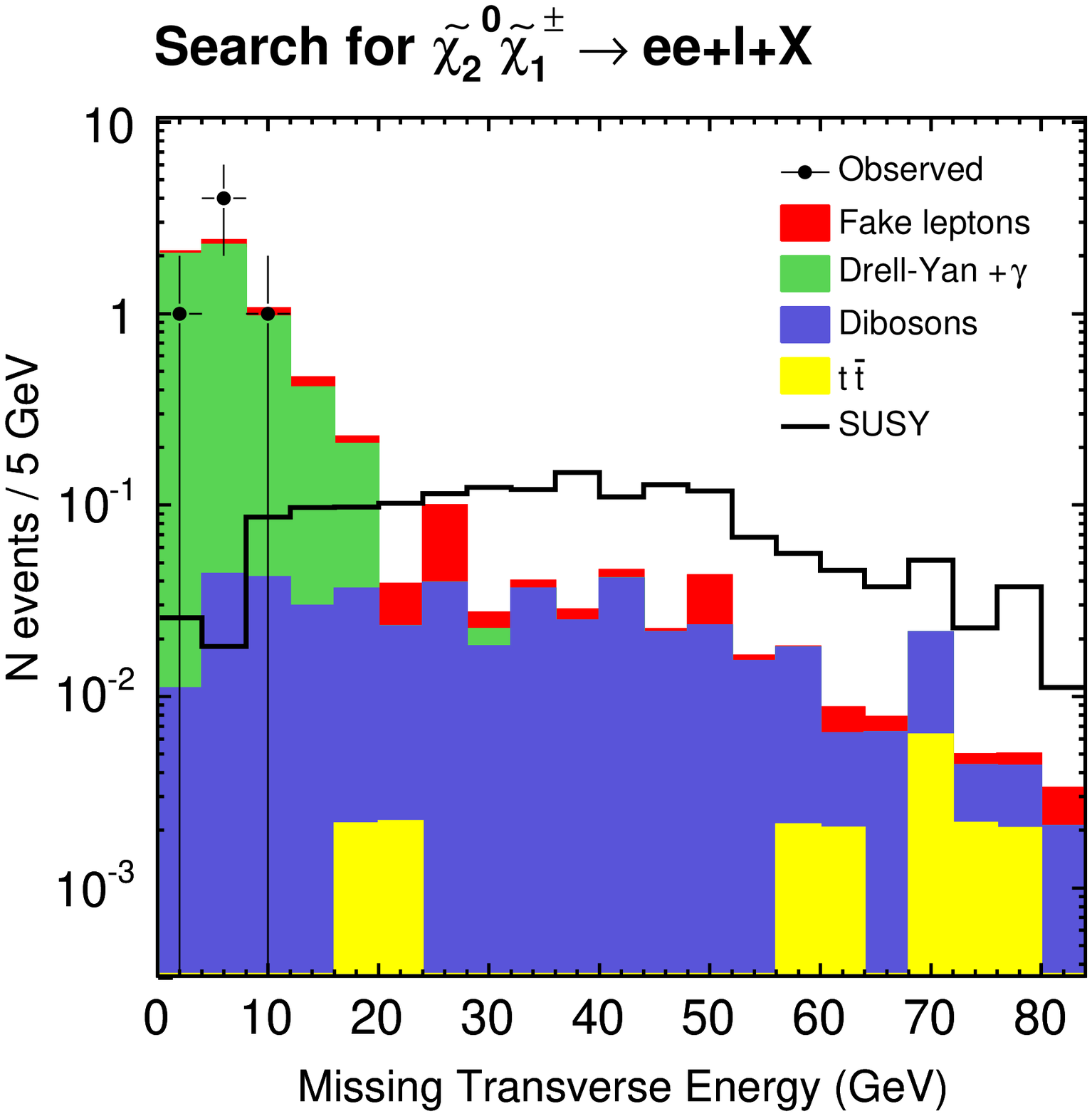}
\includegraphics[width=0.4\textwidth]{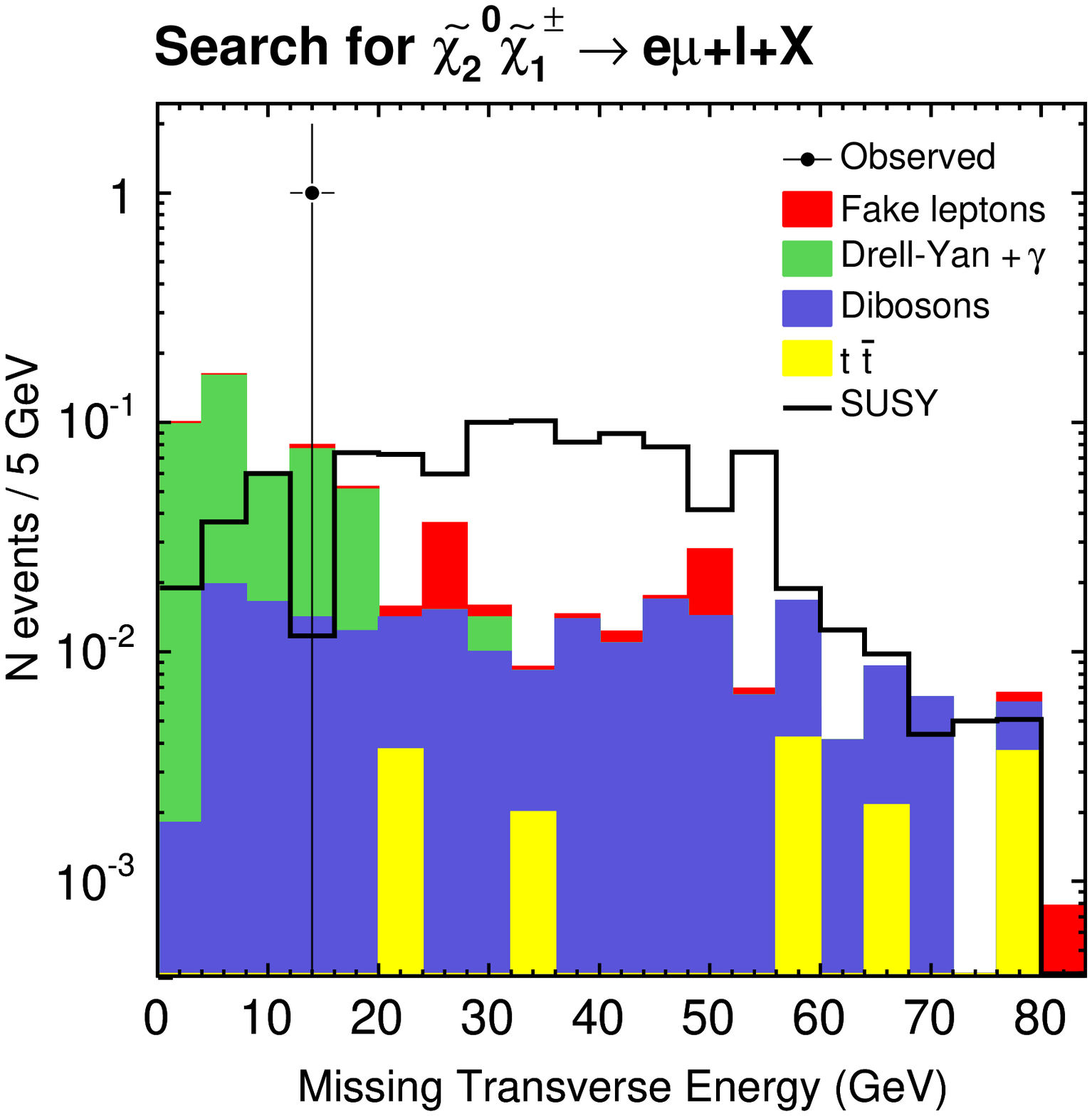} 
\caption{Missing transverse energy before the final cut of 20 GeV in $ee\ell$ and $e\mu\ell$. The SM backgrounds are stacked while the
benchmark SUSY signal is superimposed. Observed events are shown as points with error bars indicating the statistical uncertainty .}
\label{fig:electrons}\end{center}\end{figure}

\begin{figure}[ht]
\begin{center}
\includegraphics[width=0.5\textwidth]{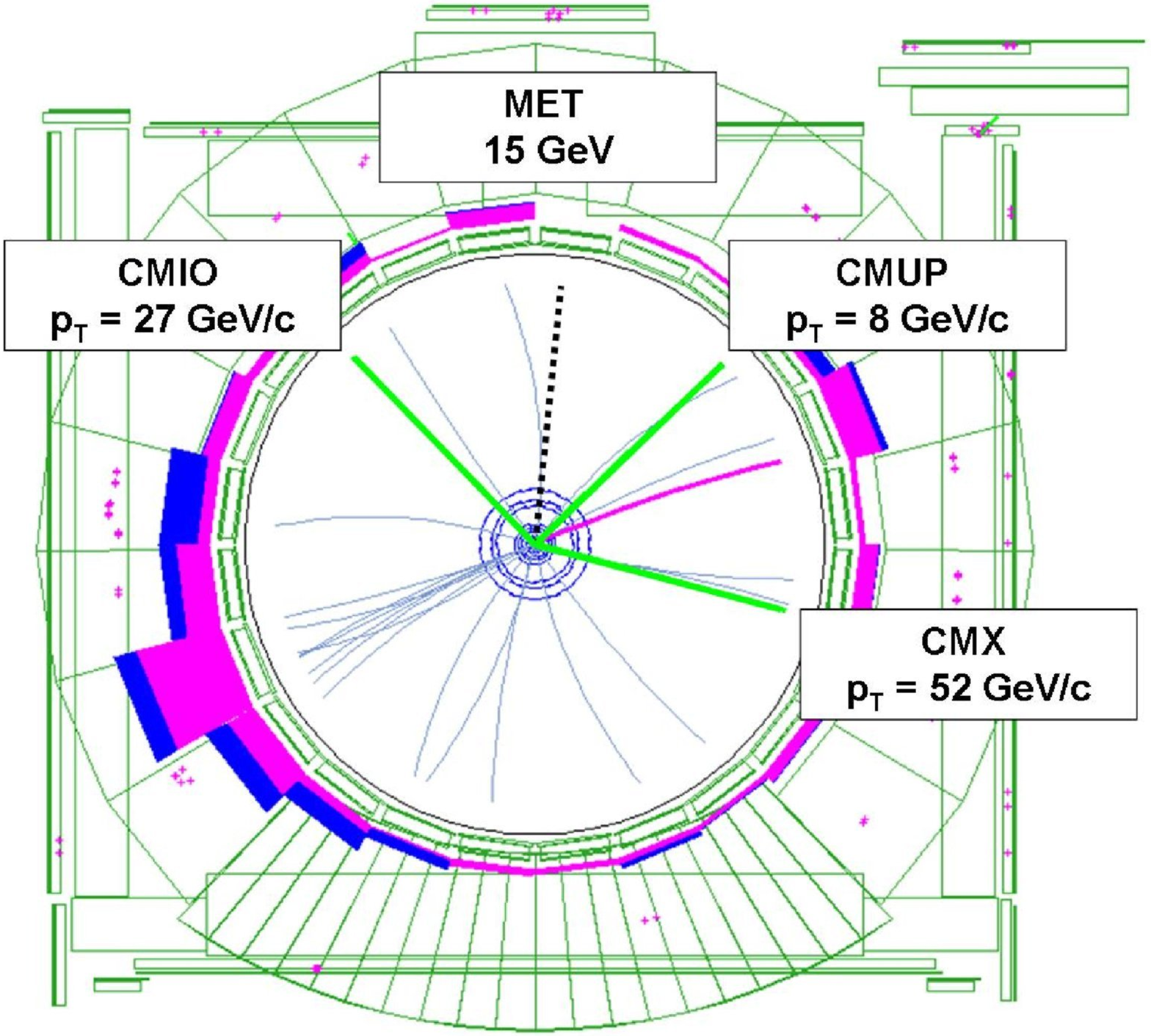}
\caption{The $r$-$\phi$ view of the $\mu\mu\mu$ candidate event in the CDF detector. Only tracks with  $p_{\rm T} \geq$ 1 GeV/$c$
are shown in the central tracking detector. The three highlighted straight tracks are labeled with the muon category and momentum. The dotted black line shows the direction of the \met\ (MET). 
The energy deposit is illustrated in the histograms around the tracking view. Innermost (light) towers show the electromagnetic energy in the calorimeter towers, outermost (dark) show the hadronic component of the energy.\label{fig::eventRPhi}}
\end{center}
\end{figure}

In Fig.~\ref{fig:muons} and Fig.~\ref{fig:electrons} 
we illustrate the \met\ in trilepton events satisfying
 the invariant mass and jet requirements. 
After applying the final cut on \met\ we observe one event and this is compatible with the SM 
predictions. The results, broken down by channels, are shown in Table 
~\ref{tab:result}.
\begin{table}[h]
\begin{center}
\caption{Summary of results for all channels. For a breakdown of individual
background components, see Table~\ref{tab:all}.}
\begin{tabular}{ccccc} \hline \hline  
& $\mu\mu \ell$& $\mu e\ell$  & $ee\ell$  & $e\mu \ell$ \\ \hline
SM Expectation &  0.6$\pm$0.2 &   0.8$\pm$0.2  & 0.6$\pm$0.1 &  0.3$\pm$0.1\\ 
Observed & 1 & 0  & 0 & 0 \\ \hline  \hline  
\label{tab:result}
\end{tabular}
\end{center}
\end{table}
In the candidate event we reconstruct three muons originating from the same primary vertex. The highest-\pt\ muon is a 
CMX muon which fired the trigger. The second muon (oppositely-charged, and at $\Delta\phi \sim 150$ degrees with respect to the leading muon) 
is a CMIO muon entering a non-fiducial part of the CMU and CMP muon chambers. The dimuon system has an invariant mass of 72 GeV/$c^2$. The third lepton selected is a CMUP muon. 
Besides the three muons, a jet originated in the same hard interaction. The missing transverse energy is just above 
the threshold of our selection  (15 GeV) with a value of 15.5 GeV. An additional 4 GeV electron candidate is  reconstructed but it comes from a different vertex. Fig. \ref{fig::eventRPhi} shows the $r$-$\phi$ view of the event in the CDF detector.

\section{Interpretation}
\label{model}
\begin{figure}
\begin{center}
\includegraphics[width=0.45\textwidth]{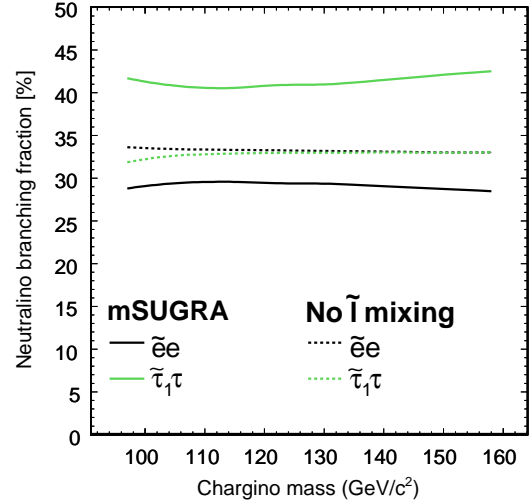}
\includegraphics[width=0.45\textwidth]{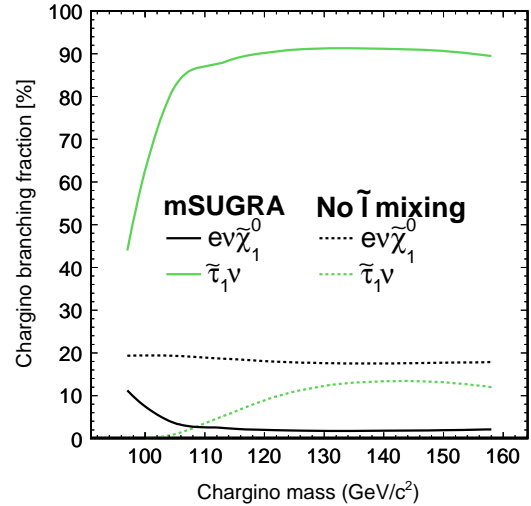}
\caption{Branching ratios for neutralino  $\tilde \chi_2^{0}$ (top) and chargino  $\tilde \chi_1^{+}$ (bottom) in the mSUGRA  
and MSSM-no-mix scenarios for different slepton flavors. The $\tilde \chi_1^{+}$ has BR($\ell\nu\tilde\chi^0_1$) 
$\approx$ BR($\bar{d} u\tilde\chi^0_1$) = BR($\bar{s}c\tilde\chi^0_1$) in both scenarios.}
\label{fig:braandb}
\end{center}
\end{figure}

\begin{figure}\begin{center}\includegraphics[width=0.46\textwidth]{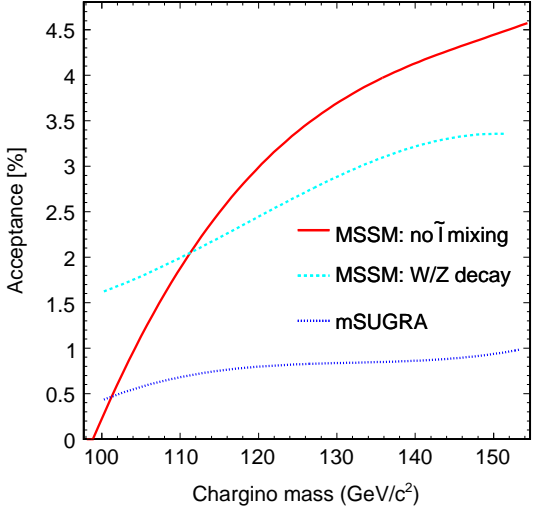}
\caption{Total acceptance of the channels for the three scenarios considered.}\label{fig:accall}\end{center}\end{figure}

We combine the four channels to obtain limits on chargino-neutralino production cross sections and masses in three SUSY models.
The calculation of the upper limit is based on the
CL$_{\mathrm{s}}$ method~\cite{tomjunk, alexread}
and incorporates the effect of the systematic uncertainties
and correlations
between channels, and between the signal and the background expectation for a
given channel.\\ 
\indent In the combination process, each simulated event and each observed event are interpreted in at most one channel. 
The overlap in the channels described above is removed by assigning shared 
SUSY signal simulated events to the analysis with the 
highest sensitivity. For a given channel the acceptance is defined as the ratio of the number of
events in the SUSY simulated sample satisfying the analysis requirements 
over the number of events where chargino $\tilde\chi_{1}^{\pm}$ and neutralino $\tilde\chi_{2}^{0}$ decay leptonically ($\tilde\chi_{1}^{\pm} \rightarrow \ell \nu \tilde\chi_{1}^{0}$, and $\tilde\chi_{2}^{0} \rightarrow \ell \ell \tilde\chi_{1}^{0}$, with $\ell = e, \mu, \tau$ ). 
In the acceptance calculation the overlap is taken into account. 
The exclusive 
background is obtained by rescaling the inclusive background 
by $\frac{A_{SUSY}^{excl}}{A_{SUSY}^{incl}}$, where $A_{SUSY}^{excl}$ ( $A_{SUSY}^{incl}$)
is the exclusive (inclusive) acceptance for the SUSY signal. This procedure
is adopted to simplify the combination with several other channels while ensuring no
double counting. We have checked that this is equivalent to the background
estimate obtained by excluding shared events within 3\%. No observed events are shared.\\
\indent We first explore the upper limit on the signal cross section times branching ratio 
in an mSUGRA scenario 
defined by the following parameters: $m_{0}$ = 60 GeV/$c^2$, $A_{0}$ = 0, $\tan\beta$ = 3, $\mu >$ 0, 
and $m_{1/2}$ varying between 162 and 230 GeV/$c^2$. These parameters were chosen 
to maximize chargino-neutralino trilepton production, $\sigma(\tilde\chi_1^{\pm}\tilde\chi_2^{0})\times BR(\tilde\chi_1^{\pm}\tilde\chi_2^{0}\rightarrow \ell\nu\tilde\chi_1^{0}\ell\ell\tilde\chi_1^{0}$).
In this scenario the two-body decays of the charginos and neutralinos into sleptons are
kinematically allowed.  \\
\indent The second model we investigate is a generic MSSM model fully defined at the electroweak scale (MSSM-W/Z).
Chargino and neutralino decays through virtual $W$ and $Z$ bosons dominate, 
resulting in three-body decays and branching
ratios similar to those of standard model $W$'s and $Z$'s~\cite{example}. In this case only the production cross section, but not the leptonic branching ratio, is dependent on the gaugino masses.\\
\indent As done in previous analyses~\cite{D0}, we also investigate a scenario in which there is no slepton mixing and the selectron, smuon, and stau 
have a degenerate mass ranging from 101 to 118 GeV/$c^2$ as $m_{1/2}$ varies between 162 and 230 GeV/$c^2$ (MSSM-no-mix). The important difference in branching ratios between scenarios mSUGRA and MSSM-no-mix is illustrated in Fig.~\ref{fig:braandb}. 
In addition to changing the mixing parameters we also increased 
the mSUGRA parameter $m_0$ to 70 GeV/$c^2$, to delay the turn-on of the $\tilde \chi^{\pm}_1 \rightarrow \tilde \nu \ell$ decay modes as the chargino mass increases.\\
\indent 
The total acceptance of the channels
described in this paper for the three scenarios is shown in Fig.~\ref{fig:accall}
 as a function of the chargino mass. 
In the MSSM-W/Z scenario the
acceptance is similar in shape but larger than the one evaluated in the
mSUGRA.  Our
sensitivity to the MSSM-W/Z is low due to the overall reduced leptonic branching ratio.
The acceptance in the mSUGRA scenario is suppressed because
of the high branching ratio into staus: the $\tilde{\tau_1}$ mass, which varies between 92 and 110 GeV/$c^2$
as $m_{1/2}$ increases from 162 to 230 GeV/$c^2$, is smaller than the first and second
generation slepton masses because of the mixing among the third-generation sleptons.
The MSSM-no-mix is a more optimistic scenario for our selection as it increases the number of electrons and muons in the final state.\\  

\begin{figure}
\begin{center}
\includegraphics[width=0.5\textwidth]{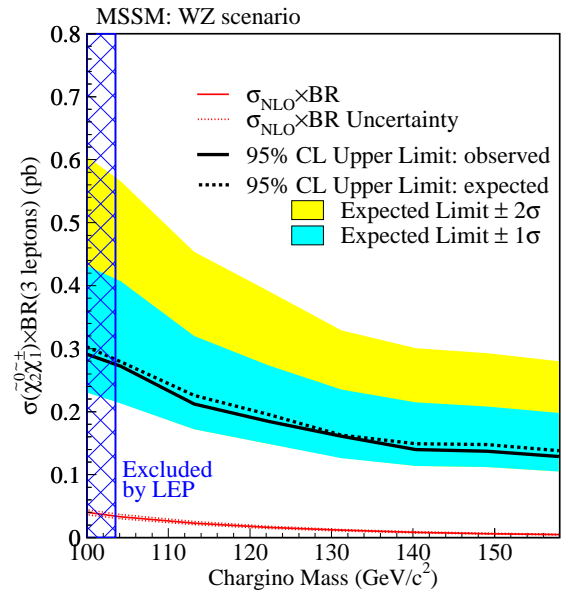}
\caption{Exclusion limits for the MSSM-W/Z scenario. The bands indicate the range of expected limits given the possible outcomes that could have been
observed if a signal were not there.}
\label{fig:limitb}
\end{center}
\end{figure}

\begin{figure}
\begin{center}
\includegraphics[width=0.5\textwidth]{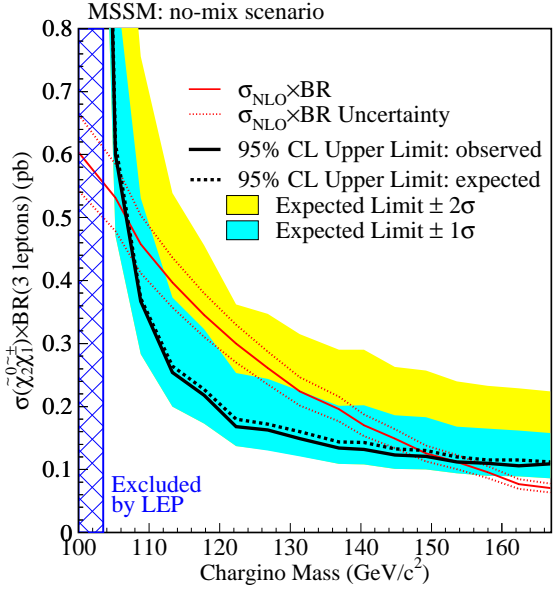}
\caption{Exclusion limits for the MSSM-no-mix scenario. The bands indicate the range of expected limits given the possible outcomes that could have been
observed if a signal were not there.}
\label{fig:limitc}
\end{center}
\end{figure}

\indent  The observed and expected limits on the cross sections 
times branching ratios are calculated at the 95$\%$ confidence level and the 
mass limits in the different scenarios are 
obtained by 
including the 
theory cross section uncertainty in the expected 
and observed limit calculation, and taking the intersection between those 
and the central theory curve. 
\begin{figure}
\begin{center}
\includegraphics[width=0.5\textwidth]{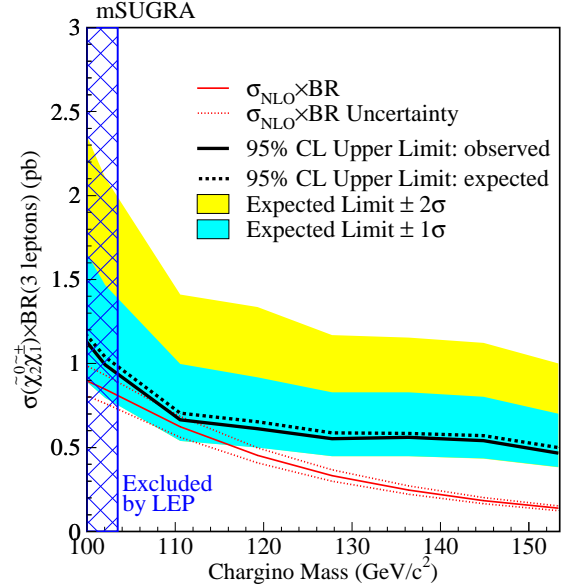}
\caption{Exclusion limits for mSUGRA scenario. The bands indicate the range
of expected limits given the possible outcomes that could have been
observed if a signal were not there.}
\label{fig:limita}
\end{center}
\end{figure}
The 95\% C.L. limits for the mSUGRA scenario and MSSM-W/Z scenario are presented in Fig.~\ref{fig:limita}. 
and Fig.~\ref{fig:limitb}.
The analyses are not sensitive to chargino and neutralino production in these models. 
For the MSSM-no-mix scenario we extend the current chargino mass limit up to 151 GeV/$c^2$ at 95\% C.L., consistent with the expected
sensitivity of 148 GeV/$c^2$ (Fig.~\ref{fig:limitc}). 
Our analysis is not sensitive to chargino masses below $\sim$110 GeV/$c^2$. This mass range represents a transition to a region of the SUSY parameter space with three-body decays of $\tilde\chi_2^0$, giving rise to very low \pt\ leptons (on average below 2 GeV/c). \\

\subsection{Projections for the CDF combined trilepton analysis}
\begin{figure} 
\begin{center}
\includegraphics[width=0.5\textwidth]{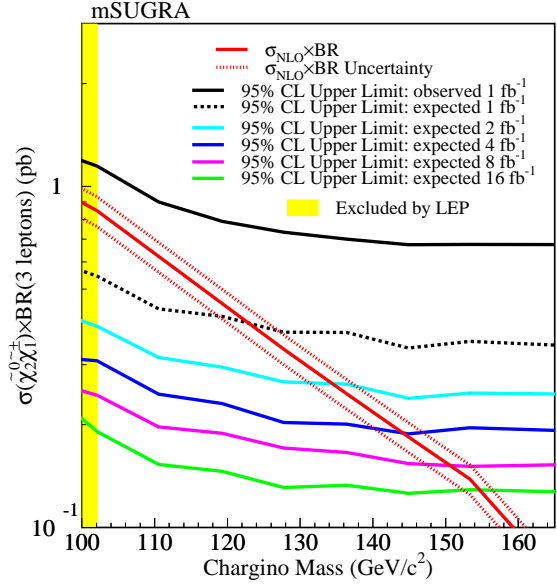}
\caption{Projection of the current expected result with increased data size for the mSUGRA scenario assuming no signal is observed. Also shown is the current observed limit. (top full line).}
\label{fig:proj1}
\end{center}
\end{figure}

\begin{figure} 
\begin{center}
\includegraphics[width=0.5\textwidth]{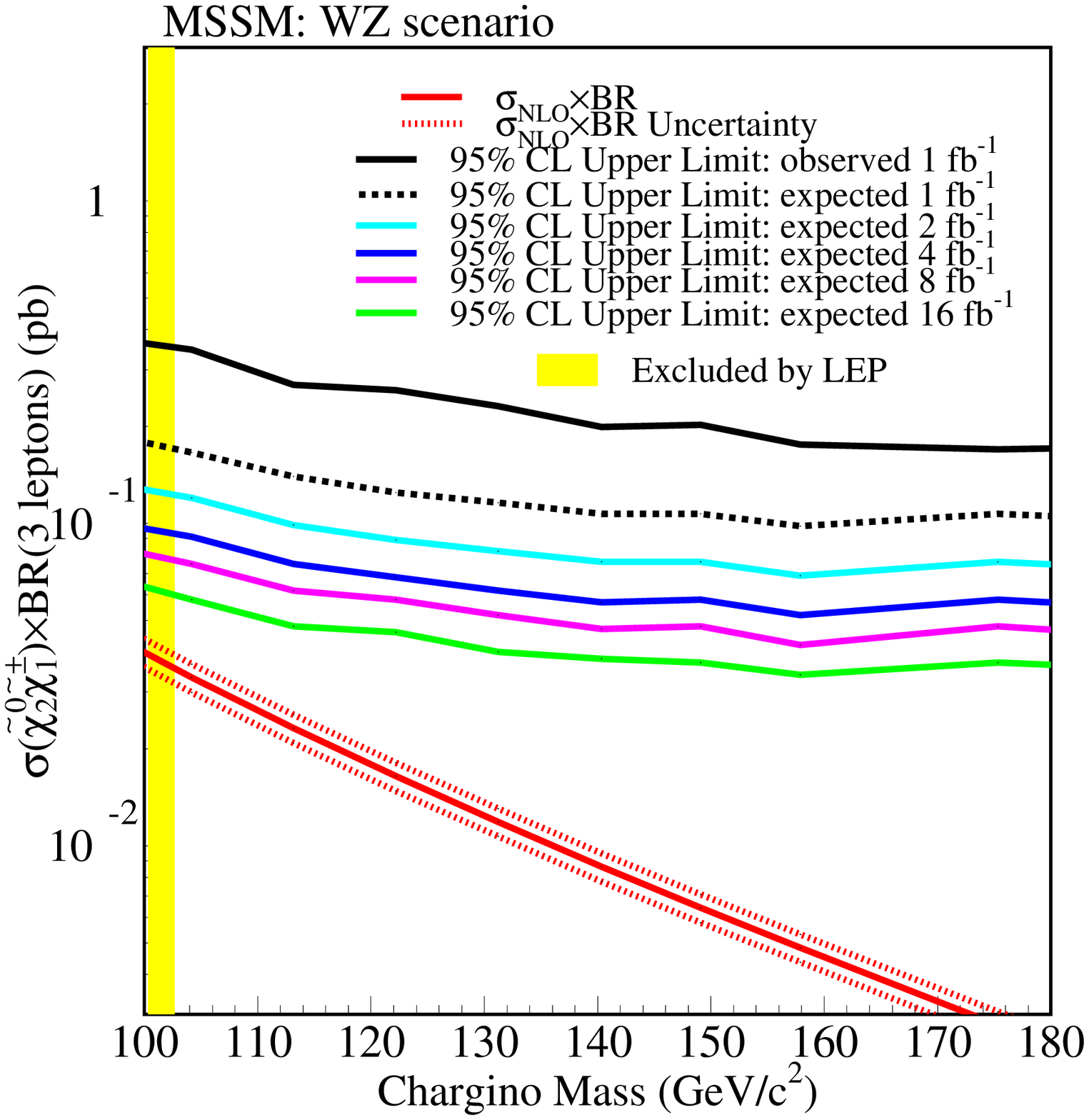}
\caption{Projection of the current expected result with increased data size for the MSSM-W/Z scenario assuming no signal is observed. Also shown is the current observed limit. (top full line).}
\label{fig:proj3}
\end{center}
\end{figure}

\begin{figure} 
\begin{center}
\includegraphics[width=0.5\textwidth]{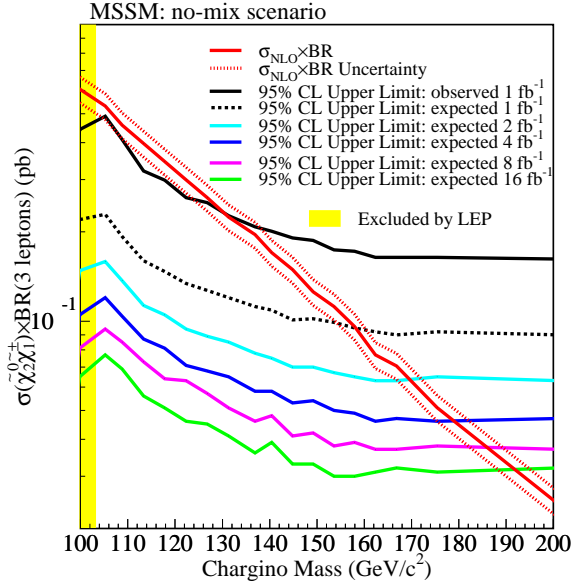}
\caption{Projection of the current expected result with increased data size for the MSSM-no-mix scenario assuming no signal is observed. Also shown is the current observed limit. (top full line).}
\label{fig:proj2}
\end{center}
\end{figure}

 The analyses presented in this paper have been combined with other CDF
searches sensitive to associated chargino-neutralino production as reported in ~\cite{prl, likesign}. 
The observed limits for the combination is less stringent than the one calculated for the high-\pt\ analysis due to slight excesses in the other channels.\\
\indent To assess the future reach at CDF, we also 
extrapolate the sensitivity of the combined analysis assuming larger data sets. Fig.~\ref{fig:proj1}, Fig.~\ref{fig:proj3}, and Fig.~\ref{fig:proj2} 
are the projected expected limits in the three models with 2, 4, 8, and 16 fb$^{-1}$ of data collected, but assuming unchanged analyses. In the plots we also assume that the 
systematic uncertainties will scale inversely with the luminosity. Using 4 fb$^{-1}$ of data, the CDF experiment has the potential to exclude chargino masses below ~140 GeV/c$^2$ and ~180 GeV/c$^2$ in the mSUGRA and MSSM-no-mix scenarios, respectively.

\section{Summary}
\label{summary}
We searched for the associated production of charginos and neutralinos in final states with one high-\pt\  electron or muon, and two additional leptons. 
The observed data counts are consistent with
the expectations from the standard model backgrounds and we set limits on the production cross section
times branching ratio. 
In the MSSM model with no slepton mixing and degenerate slepton masses, $m_0=70$ GeV/$c^2$, $\tan\beta=3$, and $\mu>0$, we set a  95\% C.L. limit on the chargino mass at 151 GeV/$c^2$.

\begin{acknowledgments}
We thank the Fermilab staff and the technical staffs of the participating institutions for their vital contributions. This work was supported by the U.S. Department of Energy and National Science Foundation; the Italian Istituto Nazionale di Fisica Nucleare; the Ministry of Education, Culture, Sports, Science and Technology of Japan; the Natural Sciences and Engineering Research Council of Canada; the National Science Council of the Republic of China; the Swiss National Science Foundation; the A.P. Sloan Foundation; the Bundesministerium f\"ur Bildung und Forschung, Germany; the Korean Science and Engineering Foundation and the Korean Research Foundation; the Science and Technology Facilities Council and the Royal Society, UK; the Institut National de Physique Nucleaire et Physique des Particules/CNRS; the Russian Foundation for Basic Research; the Comisi\'on Interministerial de Ciencia y Tecnolog\'{\i}a, Spain; the European Community's Human Potential Programme; the Slovak R\&D Agency; and the Academy of Finland.
\end{acknowledgments}

\clearpage

\clearpage

\begin{thebibliography}{99}
\bibitem{rpv} $R$-parity is defined as 
$R_{\rm P}=(-1)^{3(B-L)+2s}$, where $B$ is the baryon number, $L$ is the lepton number, and $s$ is the spin. A SM particle therefore carries $R_{\rm P}=+1$ and its SUSY partner $R_{\rm P}=-1$.
\bibitem{metdesc}  Missing transverse energy, \met, is defined as the magnitude of $-\sum_i E_T^i \hat{n_i}$, 
where $\hat n_i$ is the unit vector in the azimuthal plane that points from the interaction point to the $i^{th}$ calorimeter tower, and $E_T^i$ the transverse energy in that tower.
\bibitem{stubfootnote} A pattern of hits in a muon chamber passing certain quality criteria is defined as a ``stub''.
\bibitem{cmio} This includes cases where there is a stub in the BMU, CMU, or CMP chambers only.
\bibitem{positrons} We refer to both electrons and positrons as electrons.
\bibitem{useFake} Given the small misidentification probability per fakeable object ($\mathcal{P} \sim 10^{-3}$), it is unlikely
to find two misidentified leptons in the same event ($\mathcal{P} \sim 10^{-6}$).
\end{thebibliography}
\end{document}